\documentclass[journal,final,twocolumn]{IEEEtran}
\pdfoutput=1

\usepackage{graphicx}
\usepackage{subcaption}
\usepackage[cmex10]{amsmath}
\usepackage{amssymb}
\usepackage{amsthm}
\usepackage{cite}
\usepackage{xifthen}
\usepackage{flushend}  % for evening out columns on last page

\newboolean{isdraft}
\setboolean{isdraft}{false}
\newcommand{\ifdraft}[2]{\ifthenelse{\boolean{isdraft}}{#1}{#2}}

\newcommand{\correction}[1]{#1}

\usepackage{mathdefs}

\newcommand{\pbar}{\bar{p}}
\newcommand{\ubar}{\bar{u}}

\newcommand{\eps}{\epsilon}
\newcommand{\aeps}{\left|\eps\right|}
\newcommand{\beps}{\bar{\eps}}
\newcommand{\epsmax}{\eps_{\textup{max}}}

\newcommand{\SINR}{\textup{SINR}}

\newcommand{\gT}{g_{\textup{T}}}
\newcommand{\gR}{g_{\textup{R}}}
\newcommand{\gTI}{g_{\textup{T}_\textup{I}}}
\newcommand{\gRI}{g_{\textup{R}_\textup{I}}}

\newcommand{\GT}{G_{\textup{T}}}
\newcommand{\GR}{G_{\textup{R}}}
\newcommand{\GTI}{G_{\textup{T}_\textup{I}}}
\newcommand{\GRI}{G_{\textup{R}_\textup{I}}}

\newcommand{\fpar}[2]{%
  #1% the function name
  \ifthenelse{\isempty{#2}}{}{\!\left(#2\right)}% the function argument in a single string
}

\newcommand{\fcur}[2]{%
  #1% the function name
  \ifthenelse{\isempty{#2}}{}{\!\left\{#2\right\}}% the function argument in a single string
}

\newcommand{\fsqr}[2]{%
  #1% the function name
  \ifthenelse{\isempty{#2}}{}{\!\left[#2\right]}% the function argument in a single string
}

\newcommand{\fGT}[1]{\fpar{f_{\GT}}{#1}}
\newcommand{\fGR}[1]{\fpar{f_{\GR}}{#1}}
\newcommand{\fGTI}[1]{\fpar{f_{\GTI}}{#1}}
\newcommand{\fGRI}[1]{\fpar{f_{\GRI}}{#1}}

\newcommand{\feps}[1]{\fpar{f_{\aeps}}{#1}}
\newcommand{\dfeps}[1]{\fpar{f_{\aeps}'}{#1}}
\newcommand{\Feps}[1]{\fpar{F_{\aeps}}{#1}}
\newcommand{\sqFeps}[1]{\fpar{F_{\aeps}^{2}}{#1}}
\newcommand{\ddFeps}[1]{\fpar{F_{\aeps}''}{#1}}
\newcommand{\invFeps}[1]{\fpar{F_{\aeps}^{-1}}{#1}}

\newcommand{\TC}{\textup{TC}}
\newcommand{\TCo}{\TC_{\textup{o}}}
\newcommand{\TCs}{\TC_{\textup{s}}}

\newcommand{\TP}{\textup{TP}}
\newcommand{\TPo}{\TP_{\textup{o}}}
\newcommand{\TPs}{\TP_{\textup{s}}}

\newcommand{\lhs}{\emph{l.h.s.}}
\newcommand{\rhs}{\emph{r.h.s.}}

\newcommand{\ie}{\emph{i.e.,}}
\newcommand{\eg}{\emph{e.g.,}}
\newcommand{\ea}{\emph{et al.}}

\newcommand{\iid}{\emph{i.i.d.}}
\newcommand{\pdf}{\emph{p.d.f.}}
\newcommand{\cdf}{\emph{c.d.f.}}
\newcommand{\ccdf}{\emph{c.c.d.f.}}
\newcommand{\wrt}{\emph{w.r.t.}}
\newcommand{\rv}{\emph{r.v.}}
\renewcommand{\wp}{\emph{w.p.}}

\newtheorem{corollary}{Corollary}

\newtheorem{lemma}{Lemma}
\newtheorem{proposition}{Proposition}

\newtheorem{remark}{Remark}

\newcommand{\corref}[1]{Cor.~\ref{cor:#1}}

\newcommand{\lemref}[1]{Lem.~\ref{lem:#1}}
\newcommand{\prpref}[1]{Prop.~\ref{prp:#1}}

\renewcommand{\eqref}[1]{(\ref{eq:#1})}
\newcommand{\secref}[1]{\S\ref{sec:#1}}
\newcommand{\prfref}[1]{Appendix~\ref{prf:#1}}
\newcommand{\figref}[1]{Fig.~\ref{fig:#1}}

\begin{document}

\title{On the Joint Impact of Beamwidth and Orientation Error on Throughput in Directional Wireless Poisson Networks}


\author{%
  Jeffrey~Wildman,~\IEEEmembership{Member,~IEEE}, Pedro~H.~J.~Nardelli,~\IEEEmembership{Member,~IEEE},\\%
  Matti~Latva-aho,~\IEEEmembership{Senior~Member,~IEEE}, Steven~Weber,~\IEEEmembership{Senior~Member,~IEEE}%
  \thanks{This work has been funded by the National Science Foundation (CNS-1147838) and the Academy of Finland.}%
  \thanks{J. Wildman and S. Weber are with the Department of Electrical and Computer Engineering, Drexel University, Philadelphia, PA USA (email: \texttt{wildman@drexel.edu} and \texttt{sweber@ece.drexel.edu}).}%
  \thanks{P. Nardelli and M. Latva-aho are with the Center for Wireless Communications (CWC), University of Oulu, Finland (email: \texttt{nardelli@ee.oulu.fi} and \texttt{matti.latva-aho@ee.oulu.fi}).}%
  \thanks{S.~Weber is the corresponding author.}%
  %\thanks{Manuscript received April 19, 2005; revised December 27, 2012.}%
}

%\markboth{Journal of \LaTeX\ Class Files,~Vol.~11, No.~4, December~2012}%
%{Shell \MakeLowercase{\textit{et al.}}: Bare Demo of IEEEtran.cls for Journals}

\IEEEpubid{%
  \begin{minipage}[t]{0.85\textwidth}%
    \centering\copyright~2014 IEEE. 
    Personal use of this material is permitted. 
    Permission from IEEE must be obtained for all other uses, in any current or future media, including reprinting/republishing this material for advertising or promotional purposes, creating new collective works, for resale or redistribution to servers or lists, or reuse of any copyrighted component of this work in other works.%
  \end{minipage}%
}


\maketitle

\begin{abstract}
  We introduce a model for capturing the effects of beam misdirection on coverage and throughput in a directional wireless network using stochastic geometry.
  In networks employing ideal sector antennas without sidelobes, we find that concavity of the orientation error distribution is sufficient to prove monotonicity and quasi-concavity (both with respect to antenna beamwidth) of spatial throughput and transmission capacity, respectively.
  Additionally, we identify network conditions that produce opposite extremal choices in beamwidth (absolutely directed versus omni-directional) that maximize the two related throughput metrics.
  We conclude our paper with a numerical exploration of the relationship between mean orientation error, throughput-maximizing beamwidths, and maximum throughput, across radiation patterns of varied complexity.
\end{abstract}


% \fi

%\IEEEpeerreviewmaketitle

%=======================================================================================================================
\section{Introduction}\label{sec:intro}

\IEEEPARstart{I}{n} a wireless communications network, directional antennas can help increase received signal power while simultaneously reducing interference.
Antenna arrays that are steerable mechanically, electrically, or via switched-beams, can further improve networks performance by dynamically redirecting transmitted energy based on the network state.
The performance analysis of directional antennas in large scale wireless communications systems over the last few decades has made use of a variety of models, assumptions, and analytical tools.
However, much of the previous work assumes either perfect sector selection or beamsteering.

We anticipate the presence of several tradeoffs affecting network throughput and transmission capacity as antenna beamwidth decreases, stemming from sources of imperfect antenna configuration, beamforming, and selection \cite{RamRedSan2005, MilLloDav2009, YuYaoMol2006, Epp2006, WilGigEpp2006}.
Assuming constant transmitted power, decreasing beamwidth will result in a higher gain within the antenna's main beam.
As the beamwidth decreases, fewer interferers significantly affect the typical receiver, but their individual effects are stronger.
Additionally, as the beamwidth decreases, properly aligning the transmitter and receiver becomes more difficult, but when both are aligned, the desired signal strength increases.

In this work, we study these tradeoffs in an ad hoc wireless network setting, modeled by a bipolar Poisson Point Process (PPP).
We employ stochastic geometry to investigate optimal beamwidths that maximize throughput-based metrics in the presence of an assumed orientation error distribution.
We now categorize and review related work in this area.

%=======================================================================================================================
\subsection{Related Work}\label{sec:related-work}
\IEEEpubidadjcol

\begin{figure}[t!]
  \centering
  \ifdraft{%
    \includegraphics[width=0.375\columnwidth]{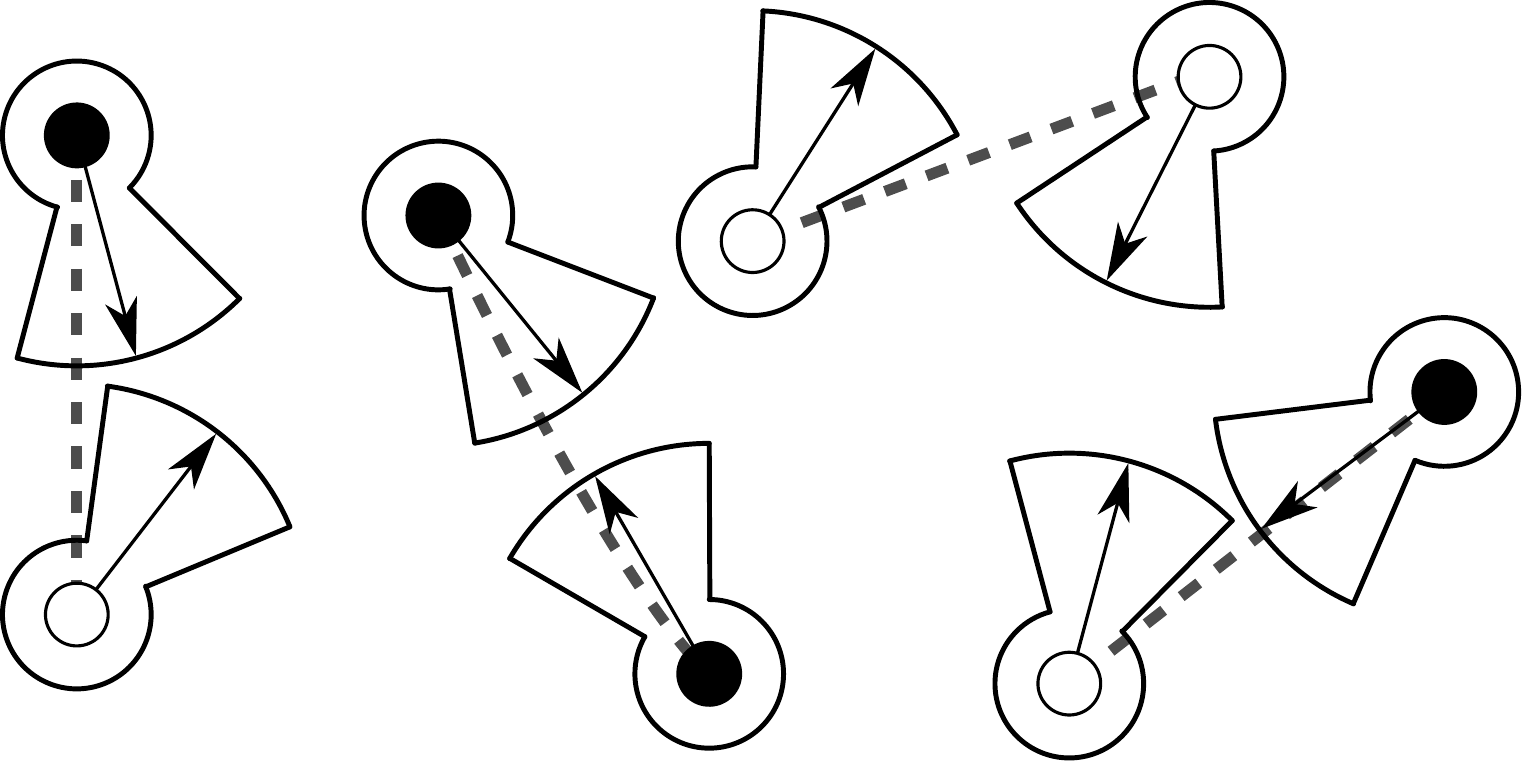}%
  }{%
    \includegraphics[width=0.75\columnwidth]{network}%
  }
  \caption{%
    A wireless network with directional antennas employed on transmitter and receiver pairs (dotted lines).
    When subjected to orientation error, transmitters and receivers may correctly direct their beam (black nodes), or they may miss their intended counterpart (white nodes).
  }
  \label{fig:network}
\end{figure}

\correction{Early works \cite{ChaCha1986, LauLeu1988} have} studied the maximum throughput and expected forward progress of ALOHA and Carrier Sense Multiple Access (CSMA) systems using the protocol model with appropriately chosen interference zones.
In particular, under Poisson ALOHA networks with antenna beamwidth \(\beta\), \correction{C.-J. Chang and J.-F. Chang \cite{ChaCha1986}} found that certain routing schemes achieve a maximum throughput gain on the order of \(\frac{1}{\beta}\).
A second set of works \cite{SpyRag2003, YiPeiShi2007, WanKonWu2010, LiZhaFan2011, CheLiuJia2013} focused on throughput capacity of directional wireless networks, extending the seminal work of Gupta and Kumar \cite{GupKum2000}.
Specifically, Yi \ea{} \cite{YiPeiShi2007} found throughput gains of \(O\left(\frac{1}{\alpha\beta}\right)\) in random networks employing the protocol interference model with transmitter and receiver beamwidths of \(\alpha\) and \(\beta\), respectively.

More recently, several works \cite{HunAndWeb2008, SinMudMad2011, WanRee2012, AkoAyaHea2012} have analyzed directional wireless networks with stochastic geometry, which offers a rich framework \cite{BacBla2010} for modeling effects such as physical interference (SINR) and fading.
Hunter \ea{} \cite{HunAndWeb2008} studied spatial diversity schemes in ad hoc networks and found that static transmit and receive beamforming with \(M\) ideal sector antennas without sidelobes yields transmission capacity gains of \(\Theta(M^2)\) over omni-directional antennas.
\correction{%
  Singh \ea{} \cite{SinMudMad2011} developed insights into medium access control design for highly directional networks by examining outage using protocol and physical interference models.
}
Wang and Reed \cite{WanRee2012} incorporated directional antennas into the analysis of coverage in multi-tier heterogeneous cellular networks.
Akoum \ea{} \cite{AkoAyaHea2012} were motivated by the rise of millimeter wave (mmWave) technology and studied achievable coverage and rates of mmWave beamsteering.

All of the works reviewed thus far assumed either perfect sector selection or perfect beamsteering.
Even the \emph{point-to-destination} scheme of \correction{C.-J. Chang and J.-F. Chang \cite{ChaCha1986}} modeled perfect orientation towards the destination, despite leading to outages when no feasible next hop falls within the transmitter's sector.
We note that antenna orientation error can affect the distribution of gains between interferers and the typical receiver.
While gain distributions have been used to study the interaction between arbitrary interferers and the typical receiver \cite{WanRee2012, AkoAyaHea2012}, we are aware of no work that incorporates orientation error into a stochastic geometry based analysis.

However, we note several works that do explicitly account for error in directional wireless networks \cite{ShePea2005, LiYaoYu2006, YuYaoMol2006, VakFriRoy2011}.
Specifically, the effect of beam-pointing error can be associated with an averaged, normalized radiation pattern with a wider main lobe and higher sidelobes than the original normalized pattern without error \cite{ShePea2005, LiYaoYu2006}.
Shen and Pearson \cite{ShePea2005} investigated coupled oscillator beam-steering arrays and the effects of per-array-element phase error on beam-pointing error.
Li \ea{} \cite{LiYaoYu2006} discussed uniform linear array (ULA) beamforming error stemming from direction of arrival (DOA) estimation, spatial (or angular) spread of the transmitted signal, antenna array element perturbation, and mutual coupling of array elements.
They analyzed outage and noted a degradation in performance due to error sources of increasing magnitude.
Vakilian \ea{} \cite{VakFriRoy2011} studied the impact of DOA estimation error, angular spread, and beamwidth on bit error rate.
They noted that narrow beamwidths can exhibit higher bit error rates than wider beamwidths when subjected to a large enough DOA error.

\subsection{Contributions}

The rest of this paper is summarized as follows.
\secref{model} introduces our wireless network model.
\secref{typical-transmission} explores the success probability of a typical transmitter under arbitrary radiation patterns (\prpref{success-general}) and sectorized patterns (\corref{success-sector} and \corref{success-sector-noside}).
%Closed form expressions for the success probability are provided for idealized sector antennas both with and without sidelobes.
\secref{tp} and \secref{tc} study network metrics spatial throughput and transmission capacity, respectively.
We derive closed form expressions for spatial throughput (\prpref{tp-sector-noside}) and transmission capacity (\prpref{tc-sector-noside}) under ideal sector antennas without sidelobes and arbitrary orientation error distributions.
We find that concavity of the cumulative orientation error distribution is sufficient for both monotonicity of spatial throughput (\prpref{tp-monotone}) and unimodality of transmission capacity (\prpref{tc-unimodal}) when expressed as functions of antenna beamwidth.
%Both monotonicity and unimodality results contain implications for the maximization of their respective metric over antenna beamwidth.
A comparison of our analytical results with that of numerical results involving more complex radiation patterns is provided in \secref{results}.
We conclude our work and outline future avenues of research in \secref{conclusions}.
Finally, for clarity, long proofs are presented in the Appendix.

%=======================================================================================================================
\section{Model}\label{sec:model}

\correction{%
  We model a wireless network with \(\hat{\Phi} = \{(x_i,m_i)\}\), a marked, homogeneous, bipolar Poisson Point Process (PPP) of intensity \(\lambda > 0\).
  \(\hat{\Phi}\) models the placement and orientation of transmitter-receiver pairs, where the members of each pair are separated by distance parameter \(d > 0\).
  The ground set \(\{x_i\} \subset \Rbb^2\) represents the transmitter (TX) locations, while the \iid{} marks \(\{m_i\}\) (formally defined in \secref{model-marks}) encode the receiver (RX) locations and antenna orientation errors.
  For notational convenience, we will denote the resulting RX locations with \(\{y_i\} \subset \Rbb^2\), where RX \(i\) is associated (paired) with TX \(i\).
  In the following subsections, we detail the rest of our model.
}

%=======================================================================================================================
\subsection{Gain Patterns}\label{sec:model-gain-patterns}

Each TX and RX is equipped with a \(2\)-dimensional antenna gain pattern \(G:[-\pi,\pi)\rightarrow\Rbb^{+}\).
The input angle to \(G(\cdot)\) is provided relative to the antenna's boresight, or `forward' direction.
For simplicity, we will assume the gain pattern is symmetric about the boresight angle: \(G(-\theta) = G(\theta)\).
Further, we will consider parameterized antenna radiation patterns such that the total radiated power (TRP) remains constant over the parameter space; equivalently, \(\int_{-\pi}^{\pi} \frac{G(\theta)}{2\pi} \drm \theta = 1\).
%
%\begin{equation}
%  \text{TRP} \equiv \int_{-\pi}^{\pi} G(\theta)\frac{P_t}{2\pi} \drm \theta = P_t.
%\end{equation}

%=======================================================================================================================
\subsection{Antenna Orientation \& Error}\label{sec:model-antennas}

Let \(\hat{\theta}_{x,y} = \angle(y-x)\) be the angle of the vector from location \(x\) to location \(y\) relative to the positive \(x\)-axis.
Thus, for a given TX-RX pair \(i\), \(\hat{\theta}_{x_i,y_i}\) is the orientation of the RX about its paired TX.
We assume that each communication device may steer its gain pattern \(G(\cdot)\) via a simple rotation of the pattern about the device's location.
Thus, under the assumption of perfect orientation, the boresight of antennas on TX \(i\) and RX \(i\) would be aligned directly along \(\hat{\theta}_{x_i,y_i}\) and \(\hat{\theta}_{y_i,x_i}\) respectively.
However, in order to study the effect of antenna misconfiguration on network performance, we introduce additive error \(\eps_x\) into the orientation of a beam originating from location \(x\), measured relative to the perfect orientation angle.
For simplicity, we will consider error distributions symmetric about zero degrees.
Finally, let \(\theta_{x_i,y_j}\) and \(\theta_{y_j,x_i}\) be the angles between TX \(i\) and RX \(j\) relative to their respective boresight angle.
These angles are the gain input angles used to compute the gain between two communication devices, and can be expressed as:
\ifdraft{%
  \begin{align}
    \theta_{x_i,y_j} &= \hat{\theta}_{x_i,y_j} - (\hat{\theta}_{x_i,y_i} + \eps_{x_i}),
  & \theta_{y_j,x_i} &= \hat{\theta}_{y_j,x_i} - (\hat{\theta}_{y_i,x_i} + \eps_{y_j}),  \label{eq:gain-angles}
  \end{align}
}{%
  \begin{align}
    \theta_{x_i,y_j} &= \hat{\theta}_{x_i,y_j} - (\hat{\theta}_{x_i,y_i} + \eps_{x_i}) \nonumber \\
    \theta_{y_j,x_i} &= \hat{\theta}_{y_j,x_i} - (\hat{\theta}_{y_i,x_i} + \eps_{y_j}), \label{eq:gain-angles}
  \end{align}
}
and are visualized by \figref{orientation}.
Note: if \(i=j\), the gain input angles between TX-RX pair \(i\) are simply \(\theta_{x_i,y_i} = -\eps_{x_i}\) and \(\theta_{y_i,x_i} = -\eps_{y_i}\).
\begin{figure}[t!]
  \centering
  \ifdraft{%
    \includegraphics[width=0.375\columnwidth]{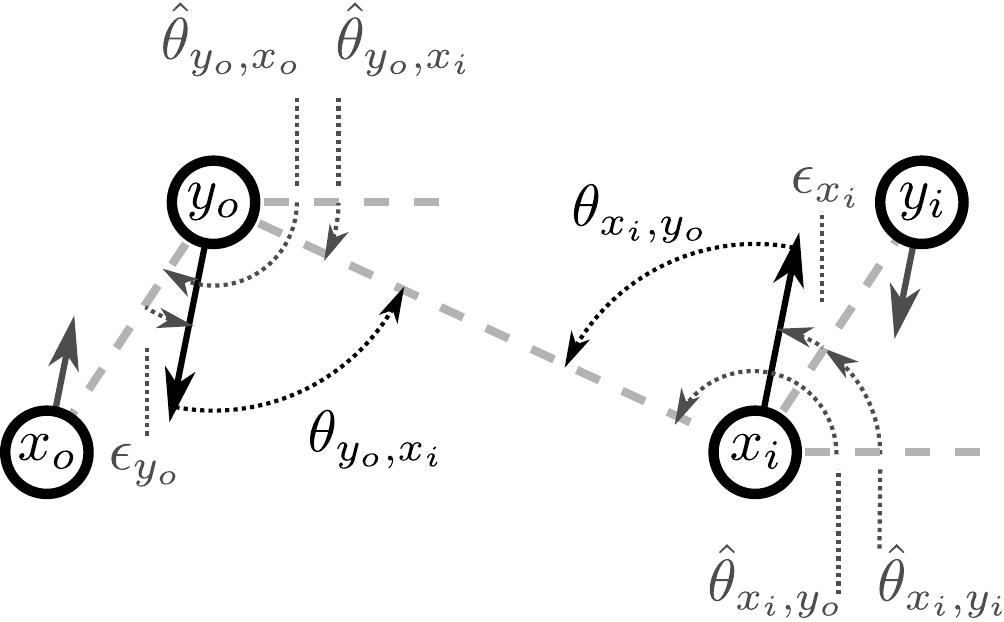}%
  }{%
    \includegraphics[width=0.75\columnwidth]{orientation}%
  }
  \caption{%
    TX/RX positions are in circles with boresight angles marked by solid arrows.
    Relevant angles are marked by gray dashed arrows, while gain input angles are black dashed arrows.
  }
  \label{fig:orientation}
\end{figure}

%=======================================================================================================================
\subsection{Marks on the Process}\label{sec:model-marks}

The marks of the process are \(\{m_i = (\hat{\theta}_{x_i,y_i},\eps_{x_i},\eps_{y_i})\}\) and consist of the following:
\begin{itemize}
  \item \(\{\hat{\theta}_{x_i,y_i} \sim [-\pi,\pi]\}\), the uniformly distributed orientation of each RX about its paired TX.
    Note: the marks encode the RX locations \(\{y_i\} \subset \Rbb^2\) via \(\{\hat{\theta}_{x_i,y_i}\}\) and \(d\),
  \item \(\{\eps_{x_i} \sim f_{\eps}:[-\epsmax,\epsmax]\rightarrow\Rbb^{+}\}\), the random orientation error of each TX's beam toward its paired RX, with zero-mean and bounded absolute error \(\epsmax \leq \pi\), and
  \item \(\{\eps_{y_i} \sim f_{\eps}:[-\epsmax,\epsmax]\rightarrow\Rbb^{+}\}\), the random orientation error of each RX's beam toward its paired TX, with zero-mean and bounded absolute error \(\epsmax \leq \pi\).
\end{itemize}

%=======================================================================================================================
\subsection{Communication Model}\label{sec:model-communication}

Finally, we model signal propagation using large-scale, distance-based pathloss with Rayleigh fading.
The signal power at RX \(j\) from TX \(i\) is given by:
\begin{equation}\label{eq:rx-power}
  P_{i,j} = P_t H_{i,j}G(\theta_{x_i,y_j})G(\theta_{y_j,x_i}) d_{i,j}^{-\alpha},
\end{equation}
where \(P_t\) is a fixed global power assignment across all transmitters, \(H_{i,j}\) is the Rayleigh fading coefficient between TX \(i\) and RX \(j\) with \(H_{i,j} \sim \text{Exp}(1)\), \(G(\theta_{x_i,y_j})\) and \(G(\theta_{y_j,x_i})\) are the gains produced by TX \(i\) and RX \(j\) respectively in the direction of each other, \(\alpha > 2\) is the large-scale pathloss constant, and \(d_{i,j}\) is the distance from TX \(i\) to RX \(j\) (note: \(d_{i,i} = d\)).

A transmission between TX-RX pair \(j\) is considered successful if the signal-to-interference-plus-noise ratio (SINR) falls above a defined SINR threshold \(\beta > 0\): \(\SINR_j = P_{j,j}/(I_j + \eta)\), where \(\eta \geq 0\) is the background noise power and \(I_j = \sum_{i \neq j} P_{i,j}\) is the sum interference power at RX \(j\).
The probability of success \(p_s\) of transmission \(j\) is then given by \(p_s = \Pbb\left\{\SINR_j \geq \beta\right\}\).
Unless otherwise noted, common parameters used to generate all figures and numerical results are \(\alpha = 3\), \(\beta = 4\), \(d = 100\) meters, \(\eta = 10^{-12}\) Watts, and \(P_t = 1\) Watt.

%=======================================================================================================================
\section{A Typical Transmission}\label{sec:typical-transmission}

In this section, we discuss the success probability of a typical TX-RX pair \(o\) with the receiver located at the origin.
This is possible due to Slivnyak's Theorem (Theorem 8.1 \cite{Hae2013}) applied to the PPP \(\hat{\Phi}\), which says that the reduced Palm distribution of \(\hat{\Phi}\) is equivalent to the original distribution of \(\hat{\Phi}\).
Here, the reduced Palm distribution of interest first conditions \(\hat{\Phi}\) on the locations of \(x_o\) and \(y_o\) and subsequently removes both points in order to provide analysis on the sum interference generated by \(\hat{\Phi}\) and observed at \(y_o\).

%=======================================================================================================================
\subsection{Induced Gain Distributions}\label{sec:typical-gain-distributions}

As Wang and Reed \cite{WanRee2012} note, the antenna gains produced between arbitrary TX/RXs in a PPP are \correction{random} variables due to their random positions.
In this work, we additionally allow the TX/RX orientations to vary independently of their positions via \(\eps_{x_i}\) and \(\eps_{y_i}\).
As a consequence, the random gains produced between the typical TX-RX pair, denoted \(\GT\) and \(\GR\), are:
\ifdraft{%
  \begin{align}
    \GT(\theta_{x_o,y_o}) &= \GT(\eps_{x_o}), \quad \eps_{x_o} \sim \feps{},
  & \GR(\theta_{y_o,x_o}) &= \GR(\eps_{y_o}), \quad \eps_{y_o} \sim \feps{}. \label{eq:GT-txfrmdists}
  \end{align}
}{%
  \begin{align}
    \GT(\theta_{x_o,y_o}) &= \GT(\eps_{x_o}), \quad \eps_{x_o} \sim \feps{}  \nonumber \\
    \GR(\theta_{y_o,x_o}) &= \GR(\eps_{y_o}), \quad \eps_{y_o} \sim \feps{}. \label{eq:GT-txfrmdists}
  \end{align}
}
Equations in \eqref{GT-txfrmdists} are due to the simplification of \eqref{gain-angles} when \(i=j=o\).
It is sufficient to consider a distribution on the absolute orientation error, \(\feps{}\), due to our assumption of a symmetric gain pattern \(G(\cdot)\).
Additionally, the random gains produced between an arbitrary TX at \(x_i\) and the typical RX at \(y_o\), denoted \(\GTI\) and \(\GRI\) are:
\ifdraft{%
  \begin{align}
    \GTI(\theta_{x_i,y_o}),& \quad \theta_{x_i,y_o} \sim [-\pi,\pi],
  & \GRI(\theta_{y_o,x_i}),& \quad \theta_{y_o,x_i} \sim [-\pi,\pi], \label{eq:GI-txfrmdists}
  \end{align}
}{%
  \begin{align}
    \GTI(\theta_{x_i,y_o}),& \quad \theta_{x_i,y_o} \sim [-\pi,\pi]  \nonumber \\
    \GRI(\theta_{y_o,x_i}),& \quad \theta_{y_o,x_i} \sim [-\pi,\pi], \label{eq:GI-txfrmdists}
  \end{align}
}
due to fact that \eqref{gain-angles} contains the sum of circular \rv{}'s, where one of the summands in each sum (\(\hat{\theta}_{x_o,y_o}\) or \(\hat{\theta}_{x_i,y_i}\)) is uniformly distributed over the circle, thus the sum is also uniformly distributed over the circle \cite{MarJup2000}.
In effect, the above four gains have been expressed independently of the geometry of the points in \(\hat{\Phi}\).
The distributions on the gains will be useful when computing the success probability of a typical transmission.

\begin{remark}
  In this paper, we restrict our attention to absolute orientation error \cdf{}s, \(\Feps{}\), that are twice differentiable and concave over bounded support \([0,\epsmax]\) with \(\epsmax \leq \pi\).
  Concave distributions are equivalently the set of monotonically decreasing distributions \(\ddFeps{x} = \dfeps{x} \leq 0, \forall x \in (0,\epsmax)\).
  This is a reasonable class of distributions to model sources of error, especially if increasingly large errors are expected to occur with decreasing likelihood.
  The assumption of twice differentiability is made to avoid distracting analytical corner cases.
  From the assumptions of concavity and the support, it follows that \(\feps{x} > 0, \forall x \in [0,\epsmax)\) and \(\Feps{x} > 0, \forall x \in (0,\epsmax]\).
  \correction{Finally, we will make use of truncated distributions (\eg{} exponential truncated to \([0,\pi]\)) and parameterize them by their mean (prior to truncation) in the text and figures.}
\end{remark}

%=======================================================================================================================
\subsection{Success Probability}\label{sec:typical-success-probability}

We begin with \prpref{success-general}, which provides a general formulation of the typical transmission success probability under arbitrary gain patterns and error distributions.
\begin{proposition}[Success of a Typical Transmission]\label{prp:success-general}
  In a network modeled by \(\hat{\Phi}\) with intensity \(\lambda > 0\), TX-RX separation distance \(d\), gain pattern \(G(\cdot)\), pathloss constant \(\alpha > 2\), background noise \(\eta \geq 0\), SINR threshold \(\beta\), orientation error distribution \(\feps{}\), and Rayleigh fading, the success probability \(p_s\) of a typical transmission can be expressed as:
  \ifdraft{%
    \begin{equation}
      p_s =      \int_{0^+}^{\infty}      \int_{0^+}^{\infty}      \exp  \left(-\lambda\pi \kappa \left(\frac{\beta}{\gT\gR}\right)^{  2/\alpha}          \Ebb\left[\GTI^{2/\alpha}\right] \Ebb\left[\GRI^{2/\alpha}\right]    d^2\right)
      e^{-\frac{\beta d^{\alpha} \eta}{P_t\gT\gR}} \fGT{\gT}\fGR{\gR}\drm \gT\drm \gR, \label{eq:success-general}
    \end{equation}
  }{%
    \begin{align}
      p_s \!&=\! \int_{0^+}^{\infty} \!\! \int_{0^+}^{\infty} \!\! \exp\!\left(-\lambda\pi \kappa \left(\frac{\beta}{\gT\gR}\right)^{\!2/\alpha} \!\!\!\! \Ebb\left[\GTI^{2/\alpha}\right] \Ebb\left[\GRI^{2/\alpha}\right] \! d^2\right) * \nonumber \\
      & \qquad \qquad \exp\!\left(-\frac{\beta d^{\alpha} \eta}{P_t\gT\gR}\right) \fGT{\gT}\fGR{\gR}\drm \gT\drm \gR, \label{eq:success-general}
    \end{align}
  }
  where \(\kappa = \Gamma(1+2/\alpha)\Gamma(1-2/\alpha)\) and both \(\Ebb[\GTI^{2/\alpha}]\) and \(\Ebb[\GRI^{2/\alpha}]\) are the \(2/\alpha\)-moments of the random gains produced between an arbitrary interferer and the typical RX.
\end{proposition}
As Wang and Reed \cite{WanRee2012} note, the joint gain distribution \(\fGTI{\gTI}\fGRI{\gRI}\) can be interpreted as a thinning probability.
The expression \(\lambda \fGTI{\gTI}\fGRI{\gRI}\) represents the intensity of transmitters from \(\hat{\Phi}\) that produce a combined gain of \(\gTI\gRI\) with the typical receiver at \(y_o\).
Further, if we ignore fading and approximate the sum interference (\(I_o\)) with the dominant interference,
\footnote{See Prop. 3.6 and Prop 4.2 from \cite{WebAnd2012}.  Under these assumptions, \(\kappa\) can be effectively treated as \(1\).}
success under each such thinned PPP would require that no interferers exist within a void zone of radius \(\left(\beta \frac{\gTI\gRI}{\gT\gR}\right)^{1/\alpha}d\) around the typical receiver.
The integral inside the exponent can be interpreted as a product of void probabilities across the independent, thinned PPPs.

With the appropriate assumptions, \prpref{success-general} can be related back to the success probability under omni-directional antennas \cite{BacBla2010}.
\begin{corollary}[Success with Omni-directional Antennas \cite{BacBla2010}]\label{cor:success-omni}
  Let an omni-directional antenna be described by \(G(\theta) = 1, \forall \theta \in [-\pi,\pi]\).
  If such antennas are employed in a network described by \prpref{success-general}, the success probability \(p_s\) of a typical TX-RX pair is: \(p_s = e^{-\lambda\pi\kappa\beta^{2/\alpha}d^2}e^{-\frac{\beta d^\alpha \eta}{P_t}}\), where \(\kappa = \Gamma(1+2/\alpha)\Gamma(1-2/\alpha)\).
  %\begin{equation}\label{eq:success-omni}
  %  p_s = e^{-\lambda\pi\kappa\beta^{2/\alpha}d^2}e^{-\frac{\beta d^\alpha \eta}{P_t}},
  %\end{equation}
\end{corollary}
\begin{IEEEproof}
  Under omni-directional antennas \(G(\theta) = 1\), the four gain distributions are equivalently: \(\GT \sim \GR \sim \GTI \sim \GRI \sim f_G(g) = \delta(g-1)\), where \correction{\(\delta(\cdot)\) is the Dirac delta function.}
  It immediately follows that \(\Ebb[\GTI^{2/\alpha}] = \Ebb[\GRI^{2/\alpha}] = 1\).
  Finally, apply these gain distributions and moments into \eqref{success-general} of \prpref{success-general}.
\end{IEEEproof}

%=======================================================================================================================
\subsection{Ideal Sectors}

Let \(G_{\textup{ideal}}(\theta)\) be an ideal sector antenna with a gain pattern defined by beamwidth \(\omega \in (0,2\pi)\), main beam gain \(g_1\), and sidelobe gain \(g_2\) with \(0 \leq g_2 < 1 < g_1\):
\begin{equation}\label{eq:pattern-sector}
  G_{\textup{ideal}}(\theta) = \begin{cases}
    g_1 = \frac{2\pi - (2\pi-\omega)g_2}{\omega} & \text{ if } |\theta| \leq \omega/2 \\
    g_2 & \text{ else}
  \end{cases}.
\end{equation}
A visualization of this pattern is provided in \figref{pattern-sector}.
Note: the TRP of \eqref{pattern-sector} is held constant at \(P_t\) over the parameter space of \(\omega\) and \(g_2\).
\lemref{sector-gain-dist} provides the four distributions and \(2/\alpha\)-moments of the sector pattern gains between the typical RX and both the typical TX and an arbitrary interfering TX.

\begin{figure}[t!]
  \begin{subfigure}[b]{0.5\linewidth}
    \centering
    \ifdraft{%
      \includegraphics[width=0.65\columnwidth]{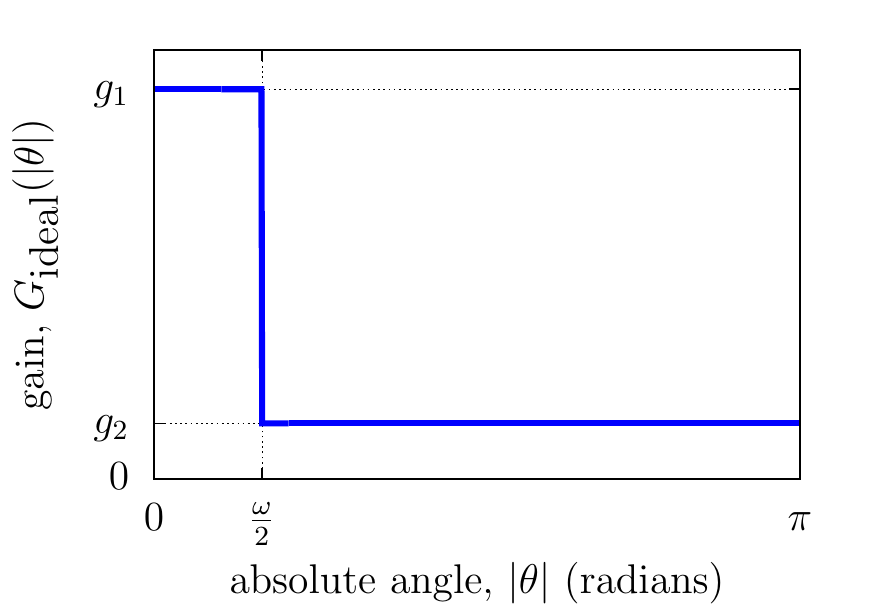}%
    }{%
      \includegraphics[width=\columnwidth]{pattern-sector-sidelobe}%
    }
  \end{subfigure}%
  \begin{subfigure}[b]{0.5\linewidth}
    \centering
    \ifdraft{%
      \includegraphics[width=0.65\columnwidth]{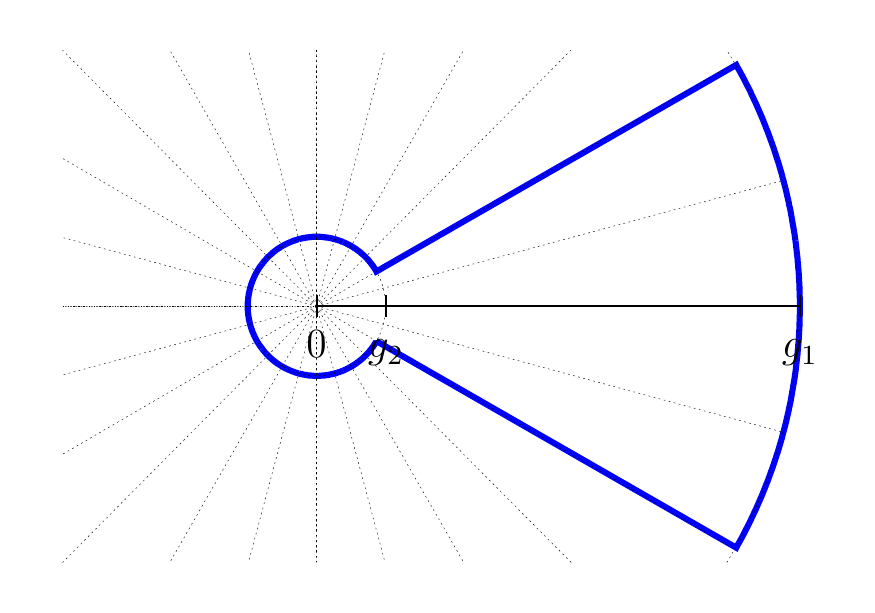}%
    }{%
      \includegraphics[width=\columnwidth]{pattern-sector-sidelobe-polar}%
    }
  \end{subfigure}
  \caption{%
    A symmetric, sector pattern with beamwidth \(\omega\), mainbeam gain \(g_1\), and sidelobe gain \(g_2\).
  }
  \label{fig:pattern-sector}
\end{figure}
\begin{lemma}[Ideal Sector Gain Distributions]\label{lem:sector-gain-dist}
  In a network modeled by \(\hat{\Phi}\) with ideal sector antennas described by \eqref{pattern-sector}, the gain distributions are given by:
  \begin{align}
    \fGT{g} = \fGR{g} = \ubar\delta(g-g_2) + u\delta(g-g_1), \\
    \fGTI{g} = \fGRI{g} = \pbar\delta(g-g_2) + p\delta(g-g_1),
  \end{align}
  where \(p = \frac{\omega}{2\pi}\), \correction{\(\pbar = 1-p\)}, \(u = \Feps{\omega/2}\) and \correction{\(\ubar = 1-u\)}.
  Further, the \(2/\alpha\)-moments of the gain distributions between an arbitrary interferer and the typical receiver are:
  \begin{equation}
    \Ebb[\GTI^{2/\alpha}] = \Ebb[\GRI^{2/\alpha}] = \pbar g_2^{2/\alpha} + pg_1^{2/\alpha}.
  \end{equation}
\end{lemma}
\begin{IEEEproof}
  Apply the ideal sector pattern \eqref{pattern-sector} to the gain distributions in \eqref{GT-txfrmdists} and \eqref{GI-txfrmdists}.
  Since \eqref{pattern-sector} produces either gains \(g_1\) and \(g_2\) over all possible input angles, the resulting gain distributions are discrete.
  The moments of the discrete gain \rv{}'s are readily computed.
\end{IEEEproof}

We note that \(p\) and \(\pbar\) can be interpreted as the main beam \emph{hit rate} and \emph{miss rate}, respectively, between interferers and the typical receiver ultimately due to their uniform orientation with respect to one another.
On the other hand, \(u\) and \(\ubar\) are the main beam hit and miss rates, respectively, between the typical TX-RX pair solely determined by their orientation errors.
\begin{corollary}[Success with Ideal Sectors]\label{cor:success-sector}
  If sectors described by \eqref{pattern-sector} with non-zero sidelobes (\(g_2 > 0\)) are employed in a network described by \prpref{success-general}, the success probability \(p_s\) of a typical TX-RX pair is:
  \ifdraft{%
    \begin{align}
      p_s &= u^2        \exp\left(-\lambda\pi \kappa\beta^{2/\alpha}d^2\left(p + \pbar\left(\frac{g_2}{g_1}\right)^{2/\alpha}\right)^2\right)                                         e^{-\frac{\beta d^\alpha \eta}{P_tg_1^2}} + \nonumber \\
      & \qquad 2u\ubar  \exp\left(-\lambda\pi \kappa\beta^{2/\alpha}d^2\left(p\left(\frac{g_1}{g_2}\right)^{1/\alpha} + \pbar\left(\frac{g_2}{g_1}\right)^{1/\alpha}\right)^2\right)  e^{-\frac{\beta d^\alpha \eta}{P_tg_1g_2}} + \nonumber \\
      & \qquad \ubar^2  \exp\left(-\lambda\pi \kappa\beta^{2/\alpha}d^2\left(p\left(\frac{g_1}{g_2}\right)^{2/\alpha} + \pbar\right)^2\right)                                         e^{-\frac{\beta d^\alpha \eta}{P_tg_2^2}}. \label{eq:success-sector}
    \end{align}
  }{%
    \begin{align}
      p_s &=    u^2     e^{       -\lambda\pi \kappa\beta^{2/\alpha}d^2\left(p + \pbar\left(\frac{g_2}{g_1}\right)^{2/\alpha}\right)^2}                                               e^{-\frac{\beta d^\alpha \eta}{P_tg_1^2}} + \nonumber \\
      & \qquad  2u\ubar e^{       -\lambda\pi \kappa\beta^{2/\alpha}d^2\left(p\left(\frac{g_1}{g_2}\right)^{1/\alpha} + \pbar\left(\frac{g_2}{g_1}\right)^{1/\alpha}\right)^2}        e^{-\frac{\beta d^\alpha \eta}{P_tg_1g_2}} + \nonumber \\
      & \qquad  \ubar^2 e^{       -\lambda\pi \kappa\beta^{2/\alpha}d^2\left(p\left(\frac{g_1}{g_2}\right)^{2/\alpha} + \pbar\right)^2}                                               e^{-\frac{\beta d^\alpha \eta}{P_tg_2^2}}. \label{eq:success-sector}
    \end{align}
  }%
  where \(\kappa\), \(p\), \(\pbar\), \(u\), and \(\ubar\) are as defined in \prpref{success-general} and \lemref{sector-gain-dist}.
\end{corollary}
\begin{IEEEproof}
  Apply the gain distributions and moments from \lemref{sector-gain-dist} to \eqref{success-general} of \prpref{success-general}.
\end{IEEEproof}

From \corref{success-sector}, the three summands in \eqref{success-sector} correspond to cases involving the typical TX and RX; \emph{i)} both the typical TX and RX \emph{hit} each other \wp{} \(u^2\), \emph{ii)} one \emph{hits} and the other \emph{misses} \wp{} \(2u\ubar\), or \emph{iii)} both \emph{miss} each other \wp{} \(\ubar^2\).
In each summand, the last exponential term relates to the transmission failure rate due to noise under Rayleigh fading, \(1-e^{-\frac{\beta d^\alpha\eta}{P_t\gT\gR}}\), and differs due to the gains between the typical TX-RX pair.
Finally, in each summand, the first exponential term contains a quadratic term in \(p\) and \(\pbar\).
This term offers some intuitive interpretations once expanded, where \(p^2\) and \(2p\pbar\) and \(\pbar^2\) represent the cases describing the hit/miss interaction between an arbitrary interferer and the typical RX and can be thought of as thinning probabilities of the interferers in \(\hat{\Phi}\).
If we ignore fading and approximate sum interference with dominant interference, the ratio of variables \(g_1\) and \(g_2\) represent adjustments to void distances/probabilities for the dominant interferer based on its hit/miss interaction with the typical RX.

Finally, if we assume perfect control over the strength of the sidelobes \(g_2=0\), we can simplify the success probability further, as given by \corref{success-sector-noside}.
\begin{corollary}[Success with Ideal Sectors \correction{without} Sidelobes]\label{cor:success-sector-noside}
  If sectors described by \eqref{pattern-sector} with zero sidelobes (\(g_2 = 0\)) are employed in a network described by \prpref{success-general}, the success probability \(p_s\) of a typical TX-RX pair is:
  \begin{equation}\label{eq:success-sector-noside}
    p_s = u^2e^{-\lambda\pi \kappa\beta^{2/\alpha}d^2 p^2}e^{-\frac{\beta d^\alpha \eta}{P_tg_1^2}}.
  \end{equation}
\end{corollary}
\begin{IEEEproof}
  Apply the gain distributions and moments from \lemref{sector-gain-dist} to \eqref{success-general}.
  Note: \(g_2 = 0\) and the lower limits of the double integration are \(0^+\).
\end{IEEEproof}

The results of \corref{success-sector-noside} offer a nice interpretation when compared to the omni-directional results of \corref{success-omni}.
First, an arbitrary interferer and the typical RX will hit each other \wp{} \(p^2\) which acts to thin the original process of interferers.
Second, within this thinned process, the typical TX-RX pair will hit each other with rate \(u^2\).

A comparison of the success probabilities established in \corref{success-omni}, \corref{success-sector}, and \corref{success-sector-noside} can be seen in \figref{success}.
The degree to which the success under ideal sector antennas with sidelobes can be approximated by that of sectors without sidelobes greatly depends on the system parameters \(\omega\), \(\Feps{}\), and \(g_2\).
Without orientation error, the three types of radiation patterns in \figref{success}, omni-directional, sectors without sidelobes, and sectors with sidelobes produce neatly ordered success curves.
Furthermore, success without sidelobes is the largest because any energy allocated to sidelobes is essentially wasted between the typical TX-RX pair in the absence of orientation error.
Finally, \(p_s \rightarrow 1\) as the network intensity \(\lambda\) (and interference) approaches zero.

\begin{figure*}[t!]
  \centering%
  \begin{subfigure}[b]{0.5\textwidth}%
    \centering%
    \includegraphics[width=\textwidth]{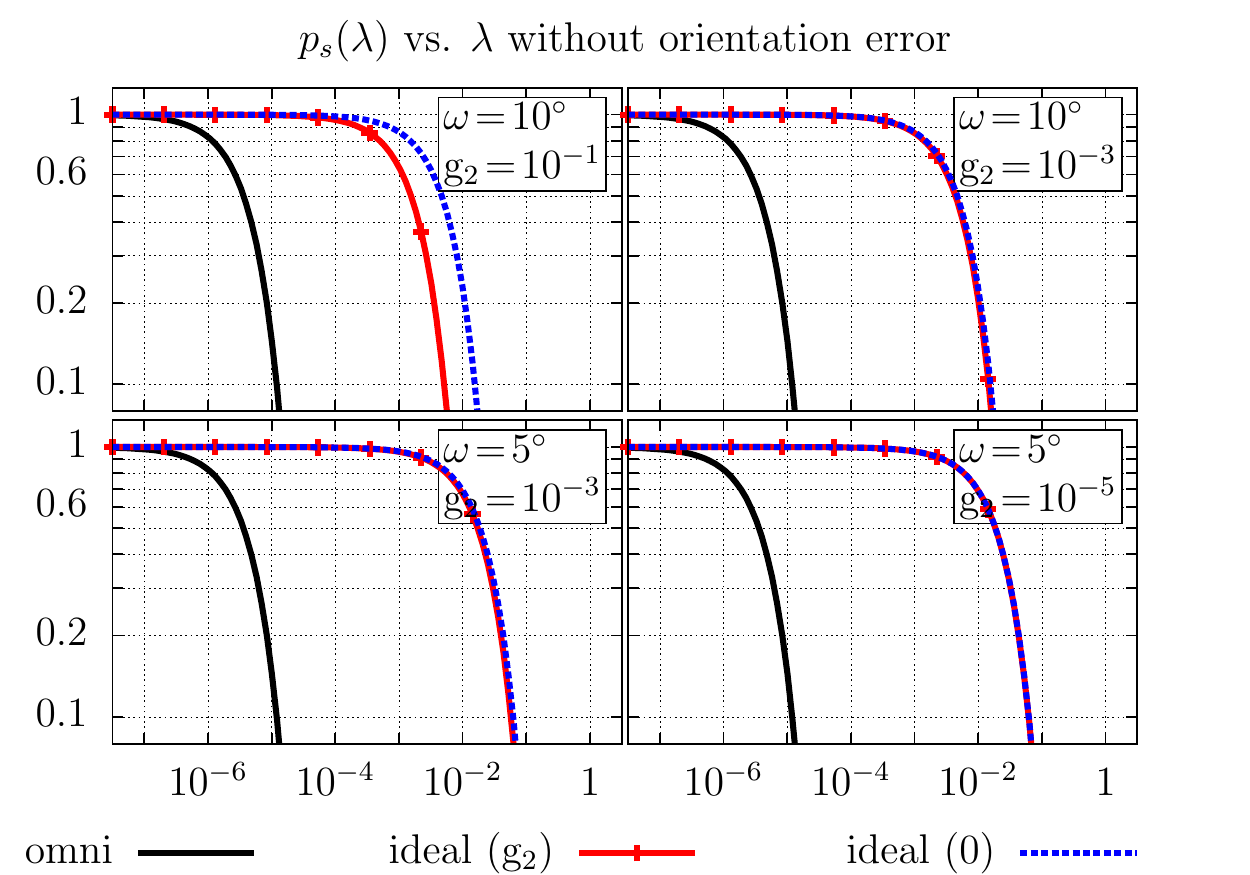}%
  \end{subfigure}%
  \begin{subfigure}[b]{0.5\textwidth}%
    \centering%
    \includegraphics[width=\textwidth]{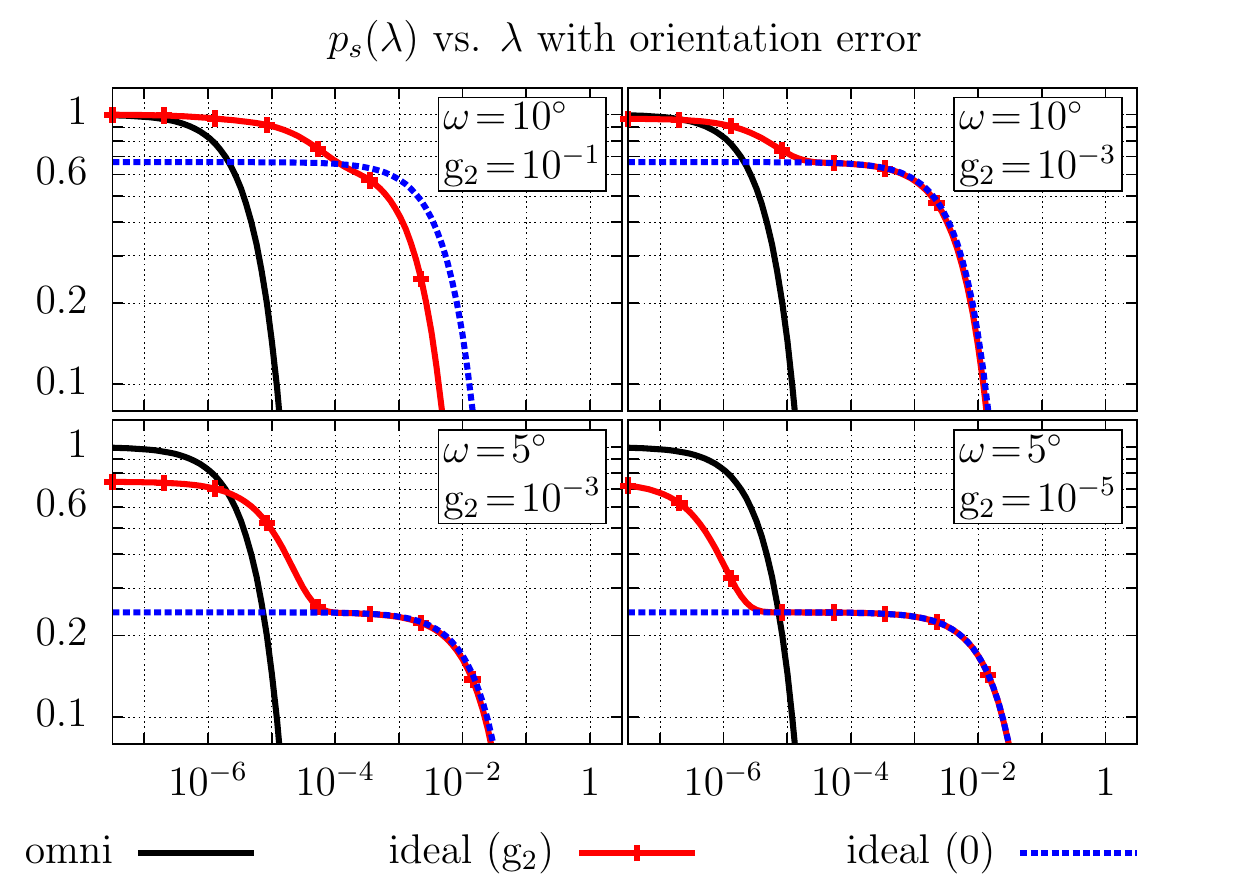}%
  \end{subfigure}%
  \caption{%
    Sample success curves \(p_s(\lambda)\) plotted against intensity of active transmitters \(\lambda\).
    %Common parameters include \(\alpha = 3\), \(\beta = 4\), \(d = 100\) meters, \(\eta = 10^{-9}\) Watts, and \(P_t=1\) Watts.
    Orientation error \(|\eps|\) is set to zero in the left set of grouped plots; in the right set of grouped plots, orientation error \(|\eps|\) is modeled using a half-normal distribution with mean \(\beps=3\) degrees.
    Curves include the cases of \textbf{omni}-directional antennas; \correction{\textbf{ideal} sector antennas with sidelobes \((g_2)\) with beamwidth \(\omega\) and sidelobe gain \(g_2\) specified in each subplot; and ideal sector antennas without sidelobes \((g_2=0)\) with the indicated beamwidth \(\omega\) specified in each subplot.}
  }
  \label{fig:success}
\end{figure*}

When orientation error is introduced, the success curves in \figref{success} are no longer guaranteed to be ordered.
In fact, the two sector types produce a crossing, where the presence of sidelobes is beneficial in low density networks, but harmful in high density networks.
The presence of sidelobes not only allows interference to be generated and received, but also permits successful communications when antennas are misaligned.
In low density networks, interference is minimal, resulting in a net benefit from sidelobes.
Regardless of the presence of error, sector antennas tend to increase a typical transmission's success for higher spatial intensities \(\lambda\) over that of omni directional antennas.
Additionally, sectors with and without sidelobes produce similar success curves when \emph{i)} the beamwidth is sufficiently larger than the mean orientation error, or \emph{ii)} the sector sidelobe is sufficiently suppressed.
As the network intensity \(\lambda\) approaches zero, success of sectors without sidelobes is upper bounded by the typical TX-RX hit rate \(u^2 = \sqFeps{\omega/2}\) which decreases as the beamwidth narrows.

%=======================================================================================================================
\section{Maximizing Spatial Throughput}\label{sec:tp}

The \emph{spatial throughput} of a network described by \prpref{success-general} is the maximum spatial intensity of successful transmissions.
Spatial throughput (\(\TP\)) is found by the maximization of \(\lambda p_s(\lambda)\) over the spatial intensity of active transmitters \(\lambda\):
\begin{equation}\label{eq:tp}
  \TP = \max_{\lambda > 0} \lambda p_s(\lambda).
\end{equation}

Spatial throughput can be achieved by an appropriate tradeoff of the intensity of active transmitters \(\lambda\) with the transmission success rate \(p_s(\lambda)\) (a monotone decreasing function of \(\lambda\)).
\prpref{tp-sector-noside} provides the optimal spatial intensity of active transmitters \(\lambda^*\) and the resulting success rate \(p_s(\lambda^*)\) that achieves \(\TP\) for networks that employ ideal sector without sidelobes.
\begin{proposition}[\(\TP\) using Sectors \correction{without} Sidelobes]\label{prp:tp-sector-noside}
  If sectors described by \eqref{pattern-sector} with zero sidelobes (\(g_2 = 0\)) are employed in a network described by \prpref{success-general}, the network's spatial throughput is \(\TPs = \lambda^*p_s(\lambda^*)\), where:
  \begin{align}
    p_s(\lambda^*) &= u^2e^{-1-\frac{\beta d^\alpha \eta}{P_t g_1^2}}, &
    \lambda^* &= \frac{1}{p^2 \pi\kappa d^2\beta^{2/\alpha}}.\label{eq:tp-sector-noside}
  \end{align}
\end{proposition}

Examining \prpref{tp-sector-noside}, the dependence of \(p_s(\lambda^*)\) and \(\lambda^*\) on \(\omega\) is such that \(p_s(\lambda^*) \rightarrow 0\) and \(\lambda^* \rightarrow \infty\) as the beamwidth is decreased \(\omega \rightarrow 0\).
It stands to reason that if \(p_s(\lambda)\) decreases slowly enough in \(\omega\), the product (spatial throughput) may be driven higher by a narrowing beamwidth.
Along these lines, we have found that spatial throughput \(\TPs\) can be increased arbitrarily, despite the presence of orientation error, under the class of concave, twice differentiable error distributions \(\Feps{}\).
This notion is formalized by \prpref{tp-monotone}.
\begin{proposition}[Concave \(\Feps{}\) Implies Monotonicity of \(\TPs\) in Beamwidth]\label{prp:tp-monotone}
  Let sectors described by \eqref{pattern-sector} with zero sidelobes (\(g_2 = 0\)) be employed in a network described by \prpref{success-general}.
  If the orientation error \cdf{} \(\Feps{}\) is concave over \([0,\pi]\), then \(\TPs\) is monotone increasing as \(\omega \rightarrow 0\).
\end{proposition}

In the proof of \prpref{tp-monotone}, we use the fact that concavity of \(\Feps{}\) implies \(\frac{\feps{x}}{\Feps{x}} \leq \frac{1}{x}\) before establishing \(\TPs\) monotonicity, producing the following nested subsets of error distributions:
\ifdraft{%
  \begin{equation}
    \{\Feps{}: \Feps{} \textrm{ concave, twice diff.} \} \subset
    \left\{\Feps{}: \frac{\feps{x}}{\Feps{x}} \leq \frac{1}{x}, \forall x > 0\right\} \subset
    \{\Feps{}: \TPs \textrm{ monotone} \}.
  \end{equation}
}{%
  \begin{align}
    &\{\Feps{}: \Feps{} \textrm{ concave, twice diff.} \} \subset \nonumber\\
    &\left\{\Feps{}: \frac{\feps{x}}{\Feps{x}} \leq \frac{1}{x}, \forall x > 0\right\} \subset \nonumber\\
    &\{\Feps{}: \TPs \textrm{ monotone} \}.
  \end{align}
}%
The ratio, \(\frac{\feps{}}{\Feps{}} = \frac{\drm}{\drm x} \log\left(\Feps{}\right)\), is also known as the logarithmic derivative of \(\Feps{}\).
We point out that truncations of a distribution \(\Feps{}\) are simply a linear scaling, and thus preserve concavity and leave the logarithmic derivative of \(\Feps{}\) unchanged.
As a result, we can classify some error distributions with infinite support (\eg{} exponential) as satisfying \prpref{tp-monotone} without having to first truncate them.
It is worth noting the possible connection with log-concavity for probability distributions, a well studied subject \cite{BagBer2005}.

A sample of \cdf{}s in the class covered by \prpref{tp-monotone} are the uniform, exponential, and normal \cdf{}s, all of which are concave, as stated by \corref{tp-monotone-dist}.
\begin{corollary}[Error Distributions with Monotone \(\TPs\)]\label{cor:tp-monotone-dist}
  If sectors described by \eqref{pattern-sector} with zero sidelobes (\(g_2 = 0\)) are employed in a network described by \prpref{success-general} with orientation error modeled by either i) uniform, ii) exponential, or iii) half-normal distributions, then \(\TPs\) is monotone decreasing in \(\omega\) over \([0,2\pi]\).
\end{corollary}
\begin{IEEEproof}
  It is enough to show that \(\Feps{x}\) is concave over \(x \in [0,\pi]\) (omitted for brevity).
  By \prpref{tp-monotone}, it follows that \(\TPs(\omega)\) is monotone decreasing in \(\omega\) over \((0,2\pi]\).
\end{IEEEproof}

\begin{remark}
  While sufficient, concavity of the error distribution \(\Feps{}\) is not necessary for throughput monotonicity.
  Consider the following error distribution on \([0,\pi]\) with a `dimple' at \((a,b)\):
  \begin{equation}\label{eq:dimple}
    \Feps{x} = \begin{cases}
      \frac{b\left(1-e^{-c_1 x}\right)}{1-e^{-c_1 a}} & 0 \leq x \leq a \\
      b + \frac{(1-b)\left(1-e^{-c_2 (x-a)}\right)}{1-e^{-c_2(\pi-a)}} & a < x \leq \pi
    \end{cases}
  \end{equation}
  \figref{dimple} displays \eqref{dimple} with parameter values such that \(\Feps{}\) is non-concave.  However, the example satisfies \(\frac{x\feps{x}}{\Feps{x}} \leq 1, \forall x > 0\) and thus yields monotone \(\TPs\).
\end{remark}
\begin{figure}[t!]
  \begin{subfigure}[b]{0.5\linewidth}
    \centering
    \ifdraft{%
      \includegraphics[width=0.65\textwidth]{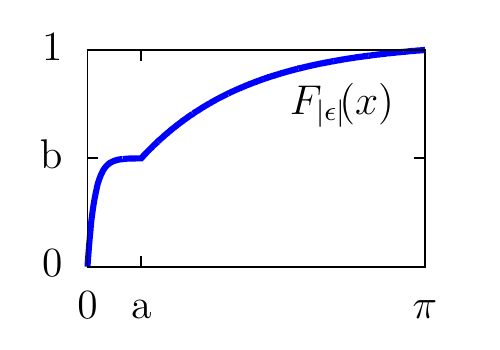}%
    }{
      \includegraphics[width=\textwidth]{dimple}%
    }
  \end{subfigure}%
  \begin{subfigure}[b]{0.5\linewidth}
    \centering
    \ifdraft{%
      \includegraphics[width=0.65\textwidth]{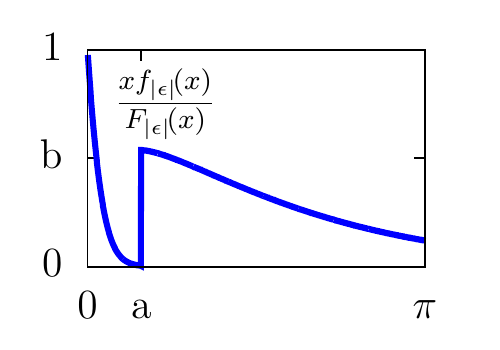}%
    }{%
      \includegraphics[width=\textwidth]{dimple2}%
    }
  \end{subfigure}
  \caption{%
    Sample distribution \({\protect\Feps{x}}\) given by \eqref{dimple} with \(a=0.5\), \(b=0.5\), \(c_1=15\), \(c_2=1\).
  }
  \label{fig:dimple}
\end{figure}

For completeness, \corref{tp-omni} provides the spatial throughput for networks employing omni-directional antennas, a well studied scenario \cite{BacBla2010, WebAnd2012}.
\begin{corollary}[\(\TP\) using Omni-directional Antennas]\label{cor:tp-omni}
  If omni-directional antennas are employed in a network described by \prpref{success-general}, the network's spatial throughput is \(\TPo = \lambda^*p_s(\lambda^*)\), where: \(p_s(\lambda^*) = e^{-1-\frac{\beta d^\alpha \eta}{P_t}}\) and \(\lambda^* = \frac{1}{\pi\kappa d^2\beta^{2/\alpha}}\).
  %\begin{align}
  %  p_s(\lambda^*) &= e^{-1-\frac{\beta d^\alpha \eta}{P_t}}, &
  %  \lambda^* &= \frac{1}{\pi\kappa d^2\beta^{2/\alpha}}.
  %\end{align}
\end{corollary}

Comparing \prpref{tp-sector-noside} and \corref{tp-omni} when background noise is negligible \((\eta \approx 0)\), the gain in \(\TP\) by the use of sectors without sidelobes over omni-directional antennas is:
\begin{equation}\label{eq:tp-gain}
  \frac{\TPs}{\TPo} = \frac{u^2}{p^2} \propto \frac{\sqFeps{\omega/2}}{\omega^2},
\end{equation}
after expanding \(g_1\), \(u\), and \(p\) in terms of \(\omega\).
As the main beam becomes narrower and stronger, \(\omega \rightarrow 0\), the \(\TP\) gain is affected by decreasing beamwidth in two opposing ways.
First, the optimal spatial intensity of active transmitters is increased by factor \(p^2 = \frac{4\pi^2}{\omega^2}\) equal to the interferer thinning probability.
Second, the success rate at this intensity decreases by factor \(u^2 = \sqFeps{\omega/2}\) equal to the hit rate of the typical TX/RX pair.
In the absence of orientation error (\(u=1\)), the use of sectorized transmitters with omni-directional receivers produces a \(\TP\) gain of \(1/p \propto 1/\omega\), which mirrors similar throughput gains derived by \correction{C.-J. Chang and J.-F. Chang \cite{ChaCha1986}}.

%=======================================================================================================================
\section{Maximizing Transmission Capacity}\label{sec:tc}

Spatial throughput is often achieved by increasing the spatial intensity of active transmitters at the expense of the success rate of the transmissions.
The \emph{transmission capacity} (TC) of a network described by \prpref{success-general} is the maximum spatial intensity of successful transmissions subject to a maximum outage constraint \(p_e\):
\begin{equation}\label{eq:tc}
  \TC = \max_{\lambda>0,p_s(\lambda)\geq 1-p_e} \lambda p_s(\lambda).
\end{equation}
As it has been well established for transmission capacity \cite{WebAnd2012}, the monotonicity of \(p_s\) in \(\lambda\) allows us to solve this maximization by taking the inverse of \(p_s(\lambda)\), which yields the intensity of active transmitters that achieves success rate \(1-p_e\):
\begin{equation}\label{eq:tc-simplified}
  \TC = \lambda(p_e) (1-p_e).
\end{equation}

\prpref{tc-sector-noside} extends the analysis of transmission capacity to networks with orientation error and ideal sector antennas without sidelobes.
\begin{proposition}[\(\TC\) using Sectors \correction{without} Sidelobes]\label{prp:tc-sector-noside}
  If sectors described by \eqref{pattern-sector} with zero sidelobes (\(g_2 = 0\)) are employed in a network described by \prpref{success-general}, the transmission capacity subject to maximum outage \(p_e\) is given by \(\TCs = \lambda^*(1-p_e)\) where:
  \begin{equation}\label{eq:tc-sector-noside}
    \lambda^* = \frac{\log\left(\frac{u^2(1-p_{\eta,s})}{1-p_e}\right)}{p^2 \pi\kappa d^2\beta^{2/\alpha}},
  \end{equation}
  and \(p_{\eta,s} = 1-e^{-\frac{\beta d^\alpha\eta}{P_tg_1^2}}\) is the failure rate due to background noise under Rayleigh fading.
\end{proposition}

Unlike the monotonicity results obtained for spatial throughput in \secref{tp}, the additional outage constraint of transmission capacity combined with antenna orientation error prohibits \(\TCs\) from being monotone increasing with the narrowing of antenna beamwidth \(\omega \rightarrow 0 \).
Ignoring background noise \((\eta \approx 0)\) for the moment, the argument to the logarithm in \eqref{tc-sector-noside} is \(\sqFeps{\frac{\omega}{2}}/(1-p_e)\) when expanded in terms of \(\omega\).
It follows that when \(\omega < 2\invFeps{\sqrt{1-p_e}}\), the transmission capacity expression will be negative, \(\TCs < 0\).
This can be interpreted in the following manner: there is a minimum threshold for beamwidth, beyond which the typical TX/RX hitting probability \(u^2\) becomes smaller than the required success rate \(1-p_e\).
In this beamwidth regime, the transmission capacity outage constraint cannot be satisfied simply due to typical TX/RX misalignment, and transmission capacity can effectively be considered zero.

Along the lines of maximizing \(\TCs\) as a function of antenna beamwidth, we have found that transmission capacity \(\TCs\) is unimodal in \(\omega\) under the class of concave, twice differentiable error distributions \(\Feps{}\).
This notion is formalized by \prpref{tc-unimodal}.
\begin{proposition}[Concavity of \(\Feps{}\) Implies Unimodality of \(\TCs\)]\label{prp:tc-unimodal}
  Let sectors described by \eqref{pattern-sector} with zero sidelobes (\(g_2 = 0\)) be employed in a network described by \prpref{success-general} with outage constraint \(p_e > 0\).
  If the orientation error \cdf{} \(\Feps{}\) is concave over \([0,\pi]\), then there exists a unique maximizer of \(\TCs\) within \(\omega \in \left(2\invFeps{\sqrt{1-p_e}},2\epsmax\right]\).
\end{proposition}
\prpref{tc-unimodal} relies on the fact that \(\TCs\) is monotone outside of the stated domain, while being quasi-concave (unimodal) inside, producing a unique maximizer.

Following the results of \prpref{tc-unimodal}, the location of the unique maximizer \(\omega^*\) of \(\TCs\) can be expressed based on conditions on \(\feps{}\) evaluated at the \rhs{} of its support \(\omega/2=\epsmax\).
\begin{corollary}[Conditions on the Maximizing \(\omega^*\) for \(\TCs\)]\label{cor:tc-maximizer}
  Let sectors described by \eqref{pattern-sector} with zero sidelobes (\(g_2 = 0\)) be employed in a network described by \prpref{success-general} with outage constraint \(p_e > 0\).
  If the orientation error \cdf{} \(\Feps{}\) is concave over \([0,\pi]\), then the unique maximizer \(\omega^*\) of \(\TCs\) has the following properties:
  \begin{itemize}
    \item If \(\feps{\epsmax} \geq \frac{\log\left(\frac{1}{1-p_e}\right)}{\epsmax}\), then \(\omega^* = 2\epsmax\).
    \item If \(\feps{\epsmax} < \frac{\log\left(\frac{1}{1-p_e}\right)}{\epsmax}\), then \(\omega^* \in \left(2\invFeps{\sqrt{1-p_e}},2\epsmax\right)\), and \(\omega^*\) is the unique solution to the equation:
      \begin{equation}\label{eq:tc-maximizer}
        \frac{\feps{\omega/2}}{\Feps{\omega/2}} = \frac{1}{\omega/2}\log\left(\frac{\sqFeps{\omega/2}}{1-p_e}\right).
      \end{equation}
  \end{itemize}
\end{corollary}
Interestingly, \corref{tc-maximizer} indicates that \(\omega^*\) depends only on the outage constraint \(p_e\) and the error distribution \(\Feps{}\) and is independent of other system parameters \(\alpha\), \(\beta\), \(\eta\), \(d\), and \(\lambda\).

\begin{figure}[t!]
  \centering
  \ifdraft{%
    \includegraphics[width=0.5\columnwidth]{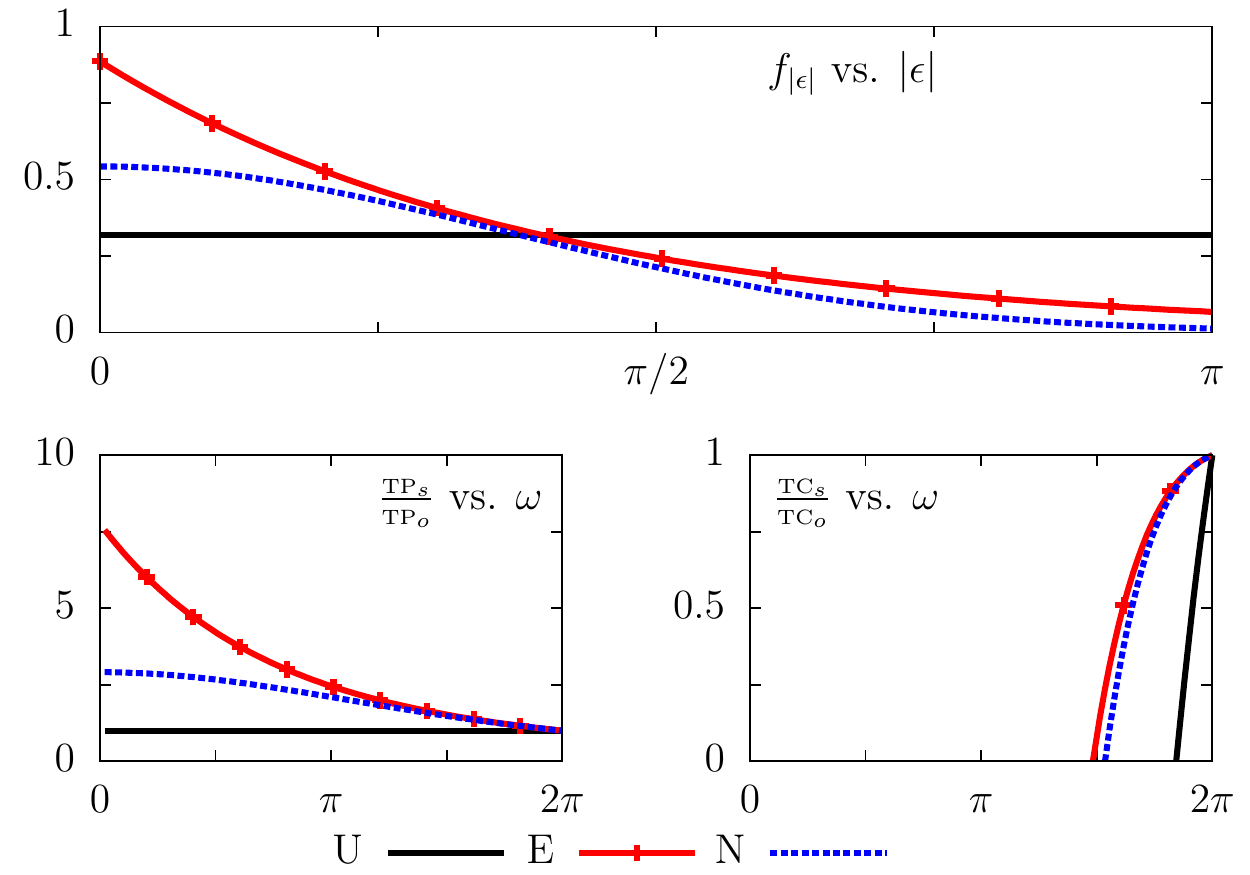}%
  }{%
    \includegraphics[width=\columnwidth]{tptc-remark}%
  }
  \caption{%
    Sample error \pdf{}s, (\textbf{U}niform, \textbf{E}xponential, and half-\textbf{N}ormal) plotted over support \(|\epsilon| \in [0,\pi]\) are shown in the top plot.
    The bottom-left and bottom-right plots display normalized spatial throughput \(\TPs/\TPo\) and normalized transmission capacity \(\TCs/\TCo\) with outage \(p_e = 0.15\) plotted against antenna beamwidth \(\omega\), respectively.
    %Common parameters include \(\alpha = 3\), \(\beta = 4\), \(d = 100\) meters, \(\eta = 10^{-9}\) Watts, \(P_t=1\) Watts, and an outage constraint of \(p_e = 0.15\).
    The \textbf{U}niform has a mean of \(\beps = 90\) degrees, while the \textbf{E}xponential and half-\textbf{N}ormal are assigned a mean of \(\beps = 70\) degrees.
    Note: the mean must be set large enough to satisfy the first condition of \corref{tc-maximizer}.
  }
  \label{fig:tptc-remark}
\end{figure}

\begin{remark}
  \corref{tc-maximizer} implies that two opposite strategies may be required to separately maximize spatial throughput and transmission capacity.
  Let \(\Feps{}\) be concave with \(\epsmax = \pi\) and let \(\feps{\epsmax} \geq \log\left(\frac{1}{1-p_e}\right)/\epsmax\).
  By concavity of \(\Feps{}\), \prpref{tp-monotone} implies that maximizing spatial throughput is done with \(\omega \rightarrow 0\).
  However, by the additional assumptions on \(\feps{\epsmax}\), \corref{tc-maximizer} implies that maximizing transmission capacity is achieved as \(\omega \rightarrow 2\epsmax = 2\pi\).
  \figref{tptc-remark} shows error distributions with parameter values such that the (normalized) spatial throughput and transmission capacity are maximized by opposite extremes of beamwidth, \(\omega \rightarrow 0\) and \(\omega \rightarrow 2\pi\).
\end{remark}

For completeness, \corref{tc-omni} provides the transmission capacity for networks employing omni-directional antennas, a well studied scenario \cite{WebAnd2012}.
\begin{corollary}[\(\TC\) using Omni-directional Antennas]\label{cor:tc-omni}
  If omni-directional antennas are employed in a network described by \prpref{success-general}, the transmission capacity subject to maximum outage \(p_e\) is \(\TCo=\lambda^*(1-p_e)\) where \(\lambda^* = \frac{\log\left(\frac{1-p_{\eta,o}}{1-p_e}\right)}{\pi\kappa d^2\beta^{2/\alpha}}\) and \(p_{\eta,o} = 1-e^{-\frac{\beta d^\alpha\eta}{P_t}}\) is the failure rate due to background noise under Rayleigh fading.
  %\begin{equation}\label{eq:tc-omni}
  %  \lambda^* = \frac{\log\left(\frac{1-p_{\eta,o}}{1-p_e}\right)}{\pi\kappa d^2\beta^{2/\alpha}},
  %\end{equation}
\end{corollary}

Comparing \prpref{tc-sector-noside} and \corref{tc-omni} when background noise is negligible \(\eta \approx 0\), we see that sector antennas without sidelobes increase the transmission capacity by a factor of:
\begin{equation}\label{eq:tc-gain}
  \frac{\TCs}{\TCo} = \frac{1}{p^2}\frac{\log\left(\frac{u^2}{1-p_e}\right)}{\log\left(\frac{1}{1-p_e}\right)} \propto \frac{1}{\omega^2}\log\left(\frac{\sqFeps{\frac{\omega}{2}}}{1-p_e}\right),
\end{equation}
after expanding \(g_1\), \(u\), and \(p\) in terms of \(\omega\).
As the main beam becomes narrower and stronger, \(\omega \rightarrow 0\), the gain in transmission capacity differs from that of spatial throughput in \eqref{tp-gain}.
The success rate is fixed at \(1-p_e\), thus the realized gain is purely a function of an adjustment to the spatial intensity of active transmitters.
While this intensity contains a similar factor \(1/p^2\) as \eqref{tp-gain}, the numerator is now \(\log\left(\sqFeps{\omega/2}/(1-p_e)\right)\), instead of simply \(\sqFeps{\omega/2}\), due to the fixed outage constraint.
In the absence of orientation error (\(u=1\)) and the employment of sectorized transmitters, the transmission capacity gain is \(1/p^2 \propto 1/\omega^2\).
When the beamwidth \(\omega\) is converted into an equivalent number of sectors \(M\) covering the circle \(\omega = \frac{2\pi}{M}\), we recover similar transmission capacity gains \(\Theta(M^2)\) derived by Hunter \ea{} \cite{HunAndWeb2008}.

%=======================================================================================================================
\section{Results}\label{sec:results}

In this section, we explore the relationship between mean orientation error, throughput maximizing beamwidths, and maximum throughput using \correction{%
  sector patterns based on those of \emph{i)} Baccelli and B\l{}aszczyszyn \cite{BacBla2010} and Akoum \ea{} \cite{AkoAyaHea2012}, and \emph{ii)} the spatial channel model used by 3GPP standards \cite{3GPP}.
  Numerical methods are used to compute the success probability of the typical transmission as well as the derived throughput metrics.
}

Based on \cite{AkoAyaHea2012}, let \(G_{\textup{trans}}(\theta)\) in \eqref{pattern-sector-transition} be an antenna gain pattern defined by \(3\)dB-beamwidth \(\omega\), main beam gain \(g_1\), and sidelobe gain \(g_2\) with \(0 \leq g_2 < g_1\) and transition width \(\gamma\):
\begin{equation}\label{eq:pattern-sector-transition}
  G_{\textup{trans}}(\theta) = \begin{cases}
    g_1 = \frac{2\pi - (2\pi - 3/2\gamma - \omega)g_2}{\omega} & \text{ if } |\theta| \leq \theta_1 \\
    g_1 - \frac{g_1}{\gamma}\left(|\theta| - \theta_1\right) & \text{ if } \theta_1 < |\theta| \leq \theta_2 \\
    \frac{2g_2}{\gamma}\left(|\theta| - \theta_2\right) & \text{ if } \theta_2 < |\theta| \leq \theta_3 \\
    g_2 & \text{ if } \theta_3 < |\theta| \leq \pi
  \end{cases},
\end{equation}
with \(\theta_1 = \omega/2-\gamma/2\), \(\theta_2 = \omega/2+\gamma/2\), and \(\theta_3 = \omega/2 + \gamma\).
In order to yield constant TRP, we place the following restrictions on the pattern's parameterization.
The beamwidth and transition width must jointly satisfy \(\omega \in (0,2\pi-2\gamma)\) and \(\gamma \in (0,\min\{\omega,\pi-\omega/2\})\) so that the full transition from main beam gain \(g_1\) to sidelobe gain \(g_2\) occurs within \(|\theta| \in [0,\pi]\).
Finally, the sidelobe gain must satisfy \(g_2 \in \left[0,1/(1-\frac{3\gamma}{4\pi})\right)\) in order for the sidelobe to be smaller than the main lobe \(g_2 < g_1\).
See \figref{pattern-sector-comparison} for a visualization of this pattern and relevant parameters.
As the transition width decreases \((\gamma \rightarrow 0)\), we recover the ideal sector pattern from \eqref{pattern-sector} with an identical beamwidth \(\omega\).
%
%Similar to \lemref{sector-gain-dist} we provide four gain distributions and relevant moments of the sector pattern with transition width \(\gamma\) in \lemref{sector-transition-gain-dist}.
%The fact that the \(2/\alpha\)-moments are closed form helps simplify the numerical integration needed to evaluate the typical transmission's probability of success provided in \prpref{success-general}.

\correction{Based on a 3GPP channel model \cite{3GPP}, let \(G_{\textup{3GPP}}(\theta)\) in \eqref{pattern-sector-3gpp} be an antenna gain pattern defined by \(3\)dB-beamwidth \(\omega\), max gain \(g_1\), and sidelobe gain \(g_2\) with \(0 \leq g_2 < g_1\):
\begin{equation}\label{eq:pattern-sector-3gpp}
  G_{\textup{3GPP}}(\theta) = \begin{cases}
    g_1 10^{-\frac{3}{10}\left(\frac{|\theta|}{\omega/2}\right)^2} & \text{ if } |\theta| \leq \theta_1 \\
    g_2 & \text{ if } \theta_1 < |\theta| \leq \pi
  \end{cases},
\end{equation}
with \(\theta_1 = \omega/2 \sqrt{10/3\log_{10}(g_1/g_2)}\) representing the angle at which the mainbeam falls off to the sidelobe level.
Note: \(g_1\) is solved for numerically to yield normalized TRP, and the parameter space of \(\omega\) and \(g_2\) is necessarily restricted so that \(\theta_1\) falls within \([0,\pi]\).}
See \figref{pattern-sector-comparison} for a visualization of this pattern and relevant parameters.

\begin{figure}[t!]
  \ifdraft{%
    \begin{subfigure}[t]{0.33\linewidth}
      \centering
      \includegraphics[width=\columnwidth]{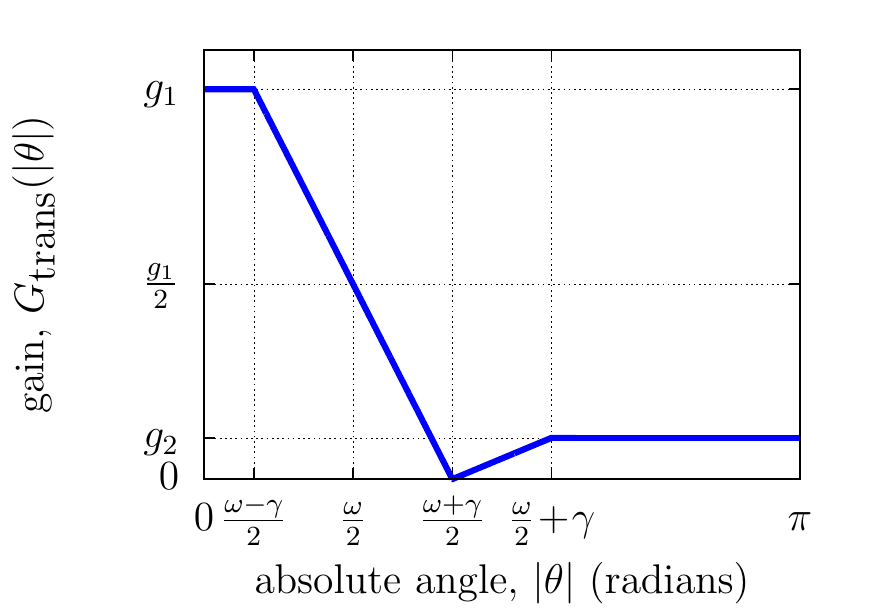}%
    \end{subfigure}%
    \begin{subfigure}[t]{0.33\linewidth}
      \centering
      \includegraphics[width=\columnwidth]{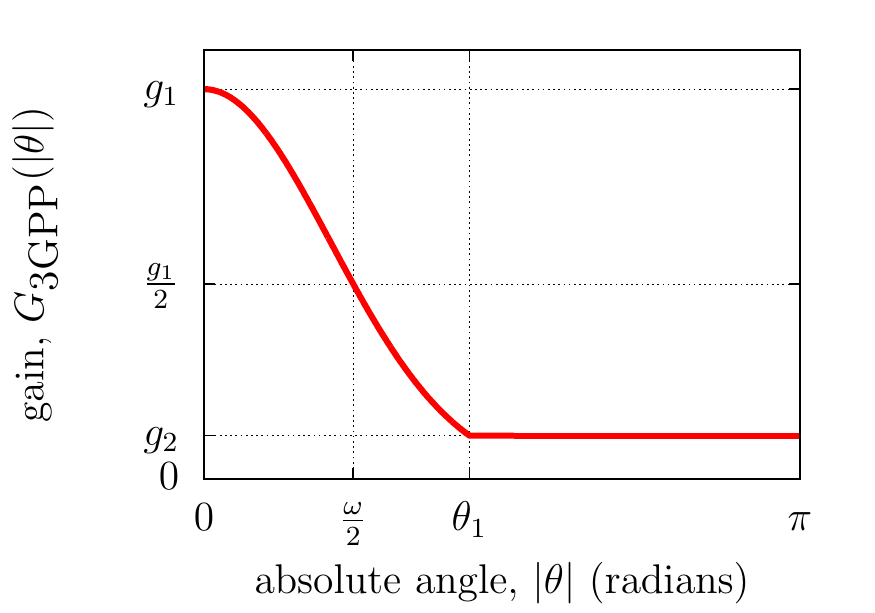}%
    \end{subfigure}%
    \begin{subfigure}[t]{0.33\linewidth}
      \centering
      \includegraphics[width=\columnwidth]{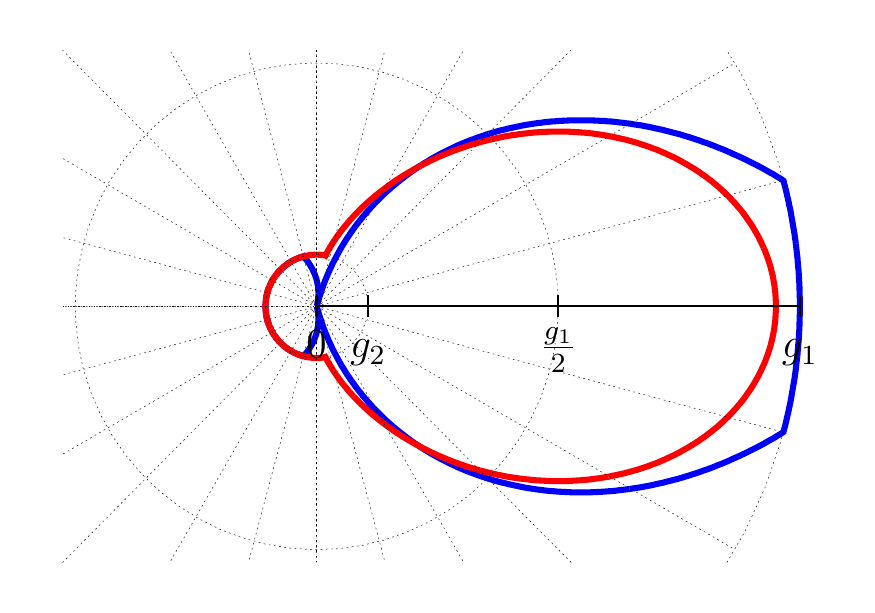}%
    \end{subfigure}
    \caption{%
      \correction{Sector patterns \eqref{pattern-sector-transition} and \eqref{pattern-sector-3gpp} (left and middle plots, respectively) with \(3\)-dB beamwidth of \(\omega\), transition width \(\gamma\), mainbeam gain \(g_1\) and sidelobe gain \(g_2\).
      Pictured in the right plot is a superimposed polar plot of both patterns for common \(g_2 = 0.35\) and \(\omega = 90\) degrees.}
    }
  }{%
    \begin{subfigure}[t]{0.5\linewidth}
      \centering
      \includegraphics[width=\textwidth]{pattern-sector-transition}%
    \end{subfigure}%
    \begin{subfigure}[t]{0.5\linewidth}
      \centering
      \includegraphics[width=\textwidth]{pattern-sector-3gpp}%
    \end{subfigure}\\
    \begin{subfigure}[t]{\linewidth}
      \centering
      \includegraphics[width=0.75\textwidth]{pattern-sector-comparison-polar}%
    \end{subfigure}
    \caption{%
      \correction{Sector patterns \eqref{pattern-sector-transition} and \eqref{pattern-sector-3gpp} (top-left and top-right plots, respectively) with \(3\)-dB beamwidth of \(\omega\), transition width \(\gamma\), mainbeam gain \(g_1\) and sidelobe gain \(g_2\).
      Pictured in the bottom plot is a superimposed polar plot of both patterns for common \(g_2 = 0.35\) and \(\omega = 90\) degrees.}
    }
  }
  \label{fig:pattern-sector-comparison}
\end{figure}

\begin{figure*}[t!]
  \ifdraft{%
    \centering
    \begin{subfigure}[b]{0.4\linewidth}
      \centering
      \includegraphics[width=0.9\textwidth]{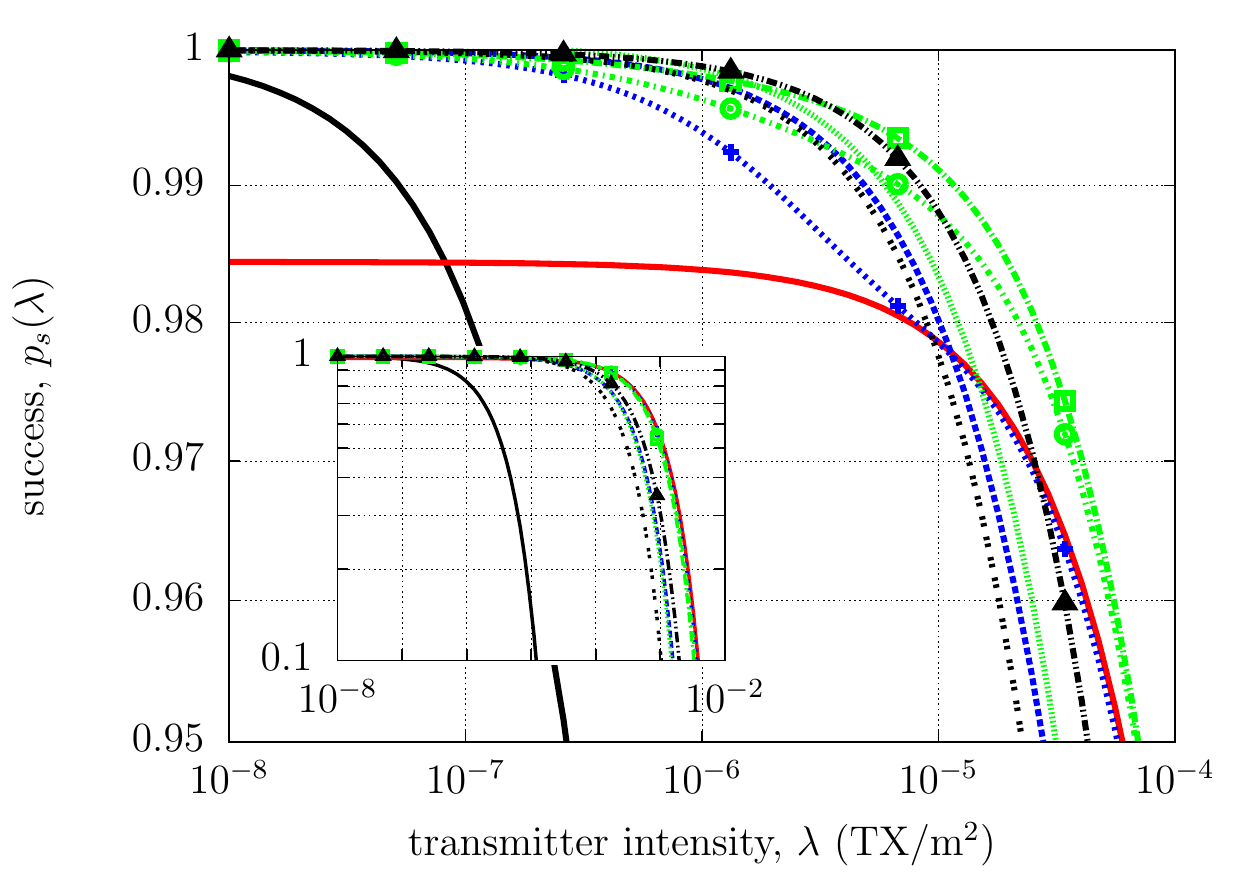}
    \end{subfigure}%
    \begin{subfigure}[b]{0.4\linewidth}
      \centering
      \includegraphics[width=0.9\textwidth]{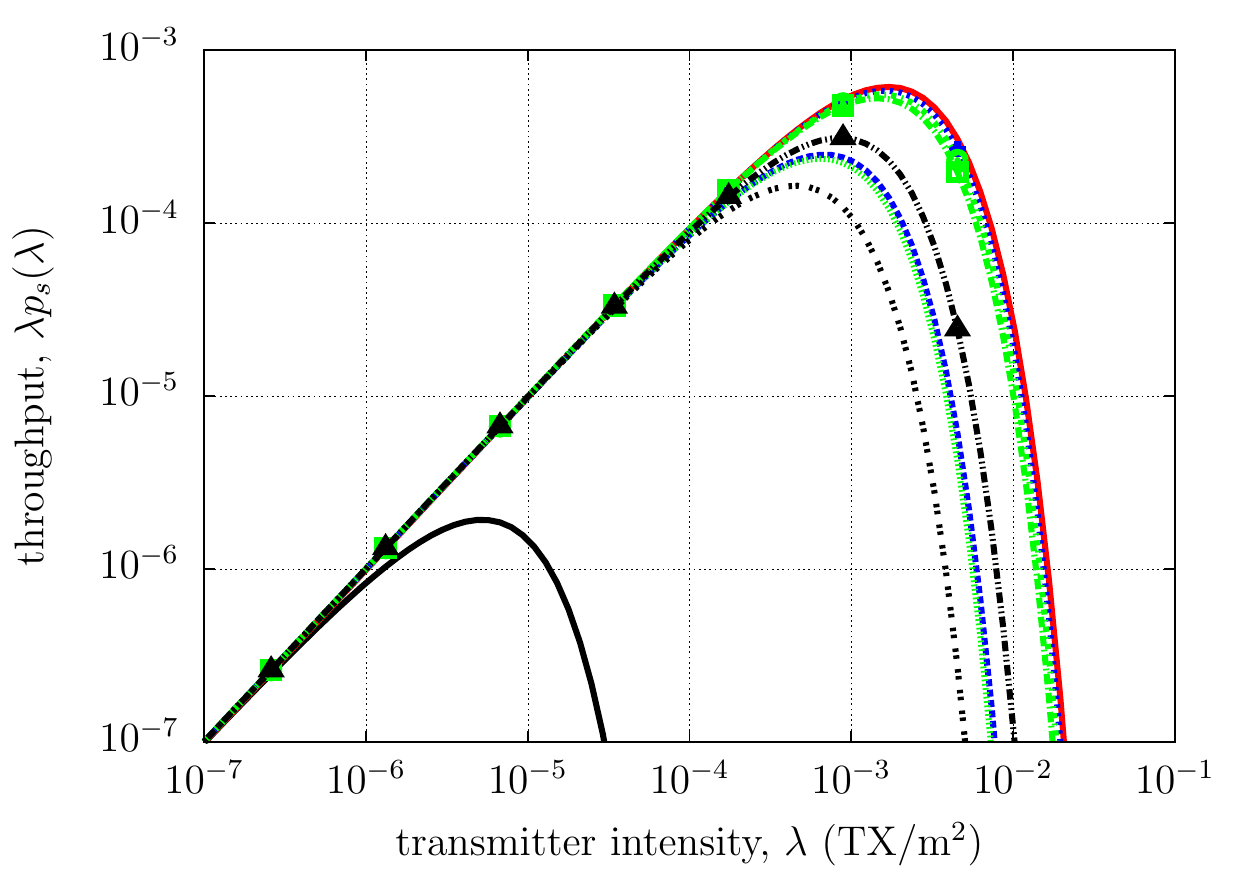}
    \end{subfigure}\\
    \begin{subfigure}[b]{0.4\linewidth}
      \centering
      \includegraphics[width=0.9\textwidth]{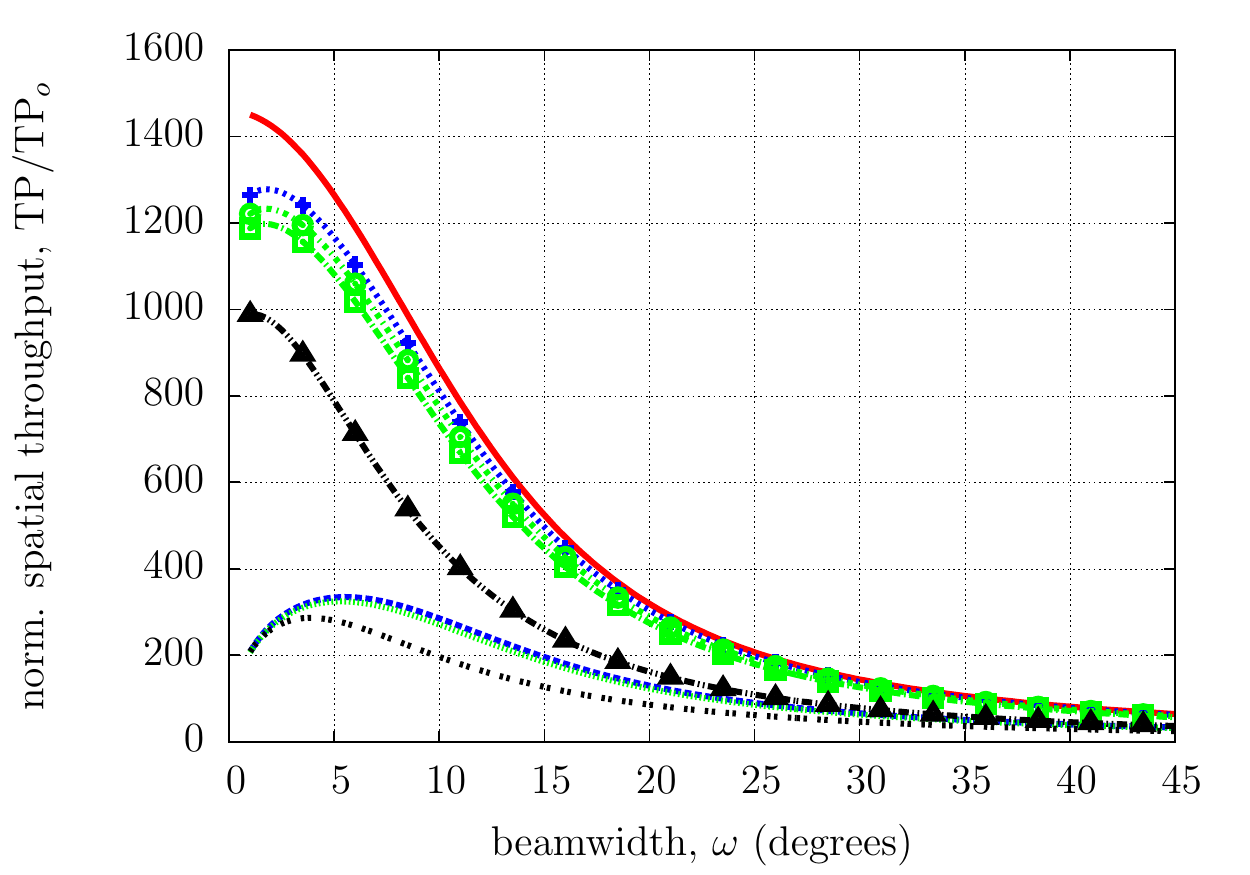}
    \end{subfigure}%
    \begin{subfigure}[b]{0.4\linewidth}
      \centering
      \includegraphics[width=0.9\textwidth]{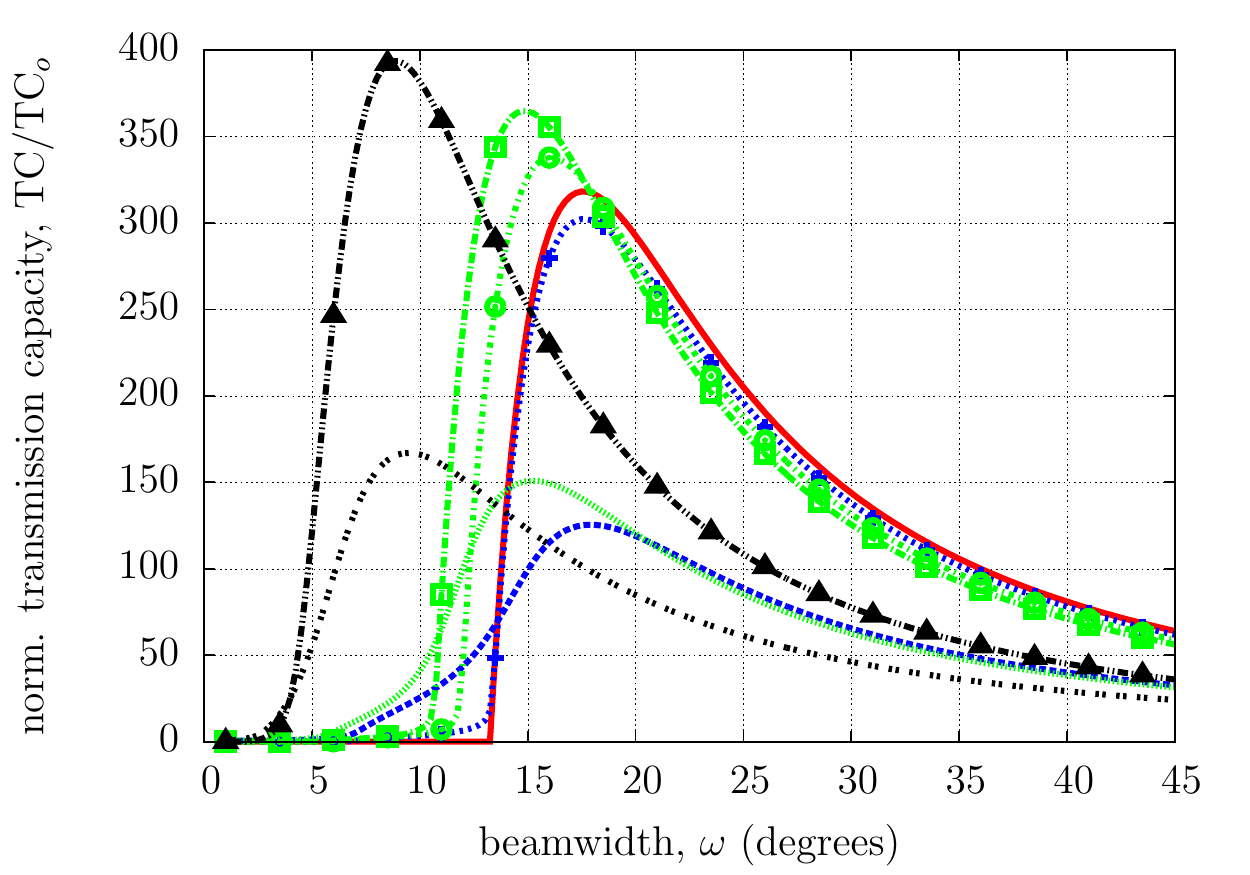}
    \end{subfigure}\\
    \begin{subfigure}[b]{0.8\linewidth}
      \centering
      \includegraphics[width=\textwidth]{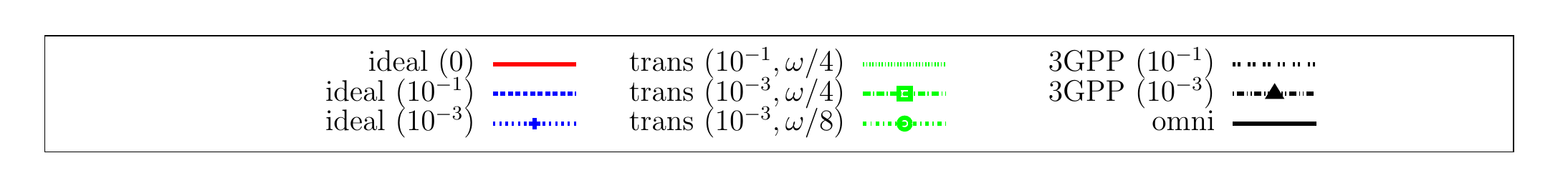}
    \end{subfigure}
  }{%
    \centering
    \begin{subfigure}[b]{0.5\linewidth}
      \centering
      \includegraphics[width=0.85\textwidth]{ps}
    \end{subfigure}%
    \begin{subfigure}[b]{0.5\linewidth}
      \centering
      \includegraphics[width=0.85\textwidth]{lpsl}
    \end{subfigure}\\
    \begin{subfigure}[b]{0.5\linewidth}
      \centering
      \includegraphics[width=0.85\textwidth]{tp}
    \end{subfigure}%
    \begin{subfigure}[b]{0.5\linewidth}
      \centering
      \includegraphics[width=0.85\textwidth]{tc}
    \end{subfigure}\\
    \begin{subfigure}[b]{\linewidth}
      \centering
      \includegraphics[width=0.9\textwidth]{legend}
    \end{subfigure}
  }
  \caption{%
    Plotted are success probability (top-left), throughput as a function of transmitter intensity \(\lambda\) (top-right), throughput maximized over \(\lambda\) (bottom-left), and outage-constrained throughput maximized over \(\lambda\) (bottom-right).
    The legend includes \textbf{omni}-directional antennas; \correction{\textbf{ideal} sector with sidelobe strength \((g_2)\); sector with sidelobe strength and \textbf{trans}ition width \((g_2,\gamma)\); and the \textbf{3GPP} sector with sidelobe strength \((g_2)\).}
    %where specific sidelobe gains and transition widths are indicated in a tuple \((g_2,\gamma)\) in the legend.
    %Default parameters include \(\alpha = 3\), \(\beta = 4\), \(d = 100\) meters, \(\eta = 10^{-9}\) Watts, \(P_t=1\) Watts, \(p_e = 0.15\), \(\beps = 3\) degrees, \(\omega = 20\) degrees, and \(g_2 = 0.1\).
    Default parameters include \(p_e = 0.15\), \(\omega = 20\) degrees, and \(g_2 = 0.1\).
    Orientation error \(|\eps|\) is a half-normal distributed \rv{} with a mean of \(\beps = 3\) degrees.
  }
  \label{fig:numerical1}
\end{figure*}

In \figref{numerical1}, we compare the network performance of several radiation patterns explored in this paper.
\correction{%
  The success rate of a typical transmission (top-left of \figref{numerical1}) is provided with inset and outset plots.
  The inset plot shows that the success curves generally tracked one another closely.
  The outset plot magnifies differences between the curves that occur at higher success rates (\(\geq 95\%\)).
  As the transition width \(\gamma\) is decreased, we observe success rates fall, perhaps due to a `broadening' of the antenna's main beam that interferes with other transmissions more than it helps cope with orientation error.
  As the sidelobe is decreased for all three sector patterns, the main beam is strengthened and we note increased success under higher TX intensities (\(\lambda \geq 10^{-5}\)).%
}%

The throughput of the network (top-right of \figref{numerical1}) shows the spatial intensity of successful transmissions plotted against TX intensity for a fixed beamwidth of \(\omega=20\) degrees and fixed mean orientation error of \(\beps = 3\) degrees.
Higher throughputs are achieved at higher TX intensities using the directional patterns over an omni-directional pattern.
At this fixed beamwidth, the sidelobe strength \(g_2\) appears to be the dominant factor (compare with transition width \(\gamma\)) in the behavior of throughput.
\correction{Throughput curves for all three patterns corresponding to stronger sidelobes (\(g_2 = 10^{-1}\)) produce lower throughput and maximizing TX intensity than smaller sidelobes (\(g_2 = 10^{-3}\)).}

Spatial throughput (throughput maximized \wrt{} \(\lambda\)) plotted against antenna beamwidth \(\omega\) (bottom-left of \figref{numerical1}), is shown for a fixed mean orientation error.
As beamwidth is decreased below \(20\) degrees, we begin to see a greater differentiation in \(\TP\) achieved by each of the evaluated radiation patterns.
%However, as either the sidelobe strength \(g_2\) or transition width \(\gamma\) decreases, the resulting \(\TP\) approaches that of the simplified sector without sidelobes.
\correction{While the analytical result of \(\TP\)-monotonicity in \prpref{tp-monotone} is reflected numerically for ideal sectors without sidelobes (\(g_2=0\)), the introduction of sidelobes and transition widths into the directional radiation pattern does not preserve spatial throughput monotonicity.}

Transmission capacity (outage-constrained throughput maximized \wrt{} \(\lambda\)) plotted against antenna beamwidth \(\omega\) (bottom-right of \figref{numerical1}) is also shown for a fixed mean orientation error.
\correction{%
  For ideal sector patterns, as sidelobes are removed and transition width is narrowed, \(\TC\) tends to increase and the maximizing \(\omega\) decreases to match the results obtained by ideal sectors without sidelobes.
  For the 3GPP pattern, we note that the \(\TC\)-maximizing beamwidth does not appear to vary with a change in sidelobe strength.
  Additionally, the \(\TC\)-unimodality in \prpref{tc-unimodal} is reflected numerically for ideal sectors without sidelobes and appears to hold experimentally for all other shown radiations patterns.
}

\correction{%
  The outage constraint \(p_e\) appears to enforce a sharp falloff in \(\TC\) as beamwidth narrows (compare with \(\TP\)).
  This is explicitly observed in the ideal sector pattern without sidelobes.
  In this case, missing the main beam, even slightly, provides no throughput benefit.
  In all patterns with small sidelobes \(g_2=10^{-3}\), the falloff in \(\TC\) is very similar with the exception of a tail on the \lhs{}.
  The \(\TC\)-maximizing beamwidth appears more sensitive to the pattern type (ideal, trans, 3GPP) while the maximum \(\TC\) appears more sensitive to the sidelobe gain \(g_2 \in \{10^{-1}, 10^{-3}\}\).
}
Ultimately, the outage constraint prohibits blind throughput maximization \wrt{} \(\lambda\) at the expense of success, thus lower throughput is realized with \(\TC\) versus \(\TP\).

\begin{figure*}[t!]
  \ifdraft{%
    \centering
    \begin{subfigure}[b]{0.4\linewidth}
      \centering
      \includegraphics[width=0.9\textwidth]{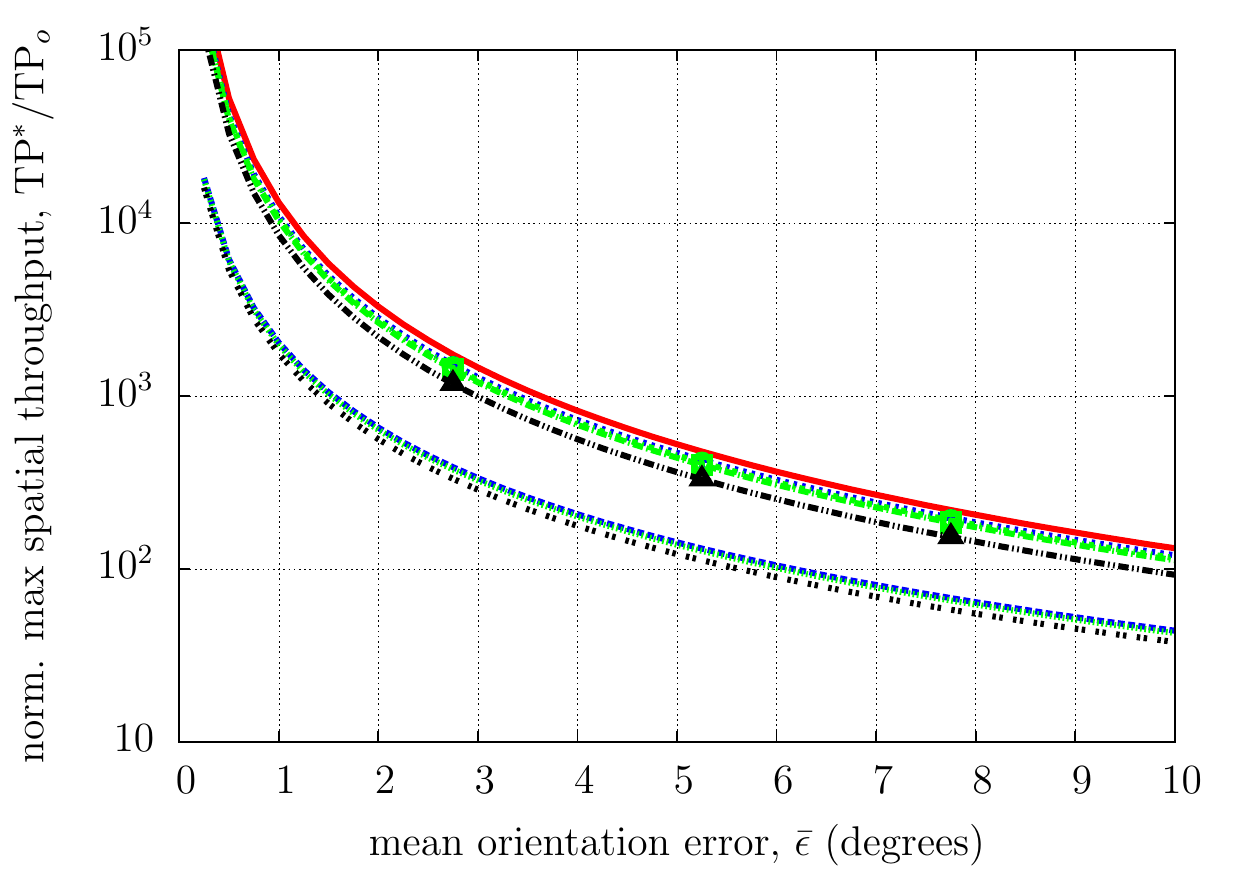}
    \end{subfigure}
    \begin{subfigure}[b]{0.4\linewidth}
      \centering
      \includegraphics[width=0.9\textwidth]{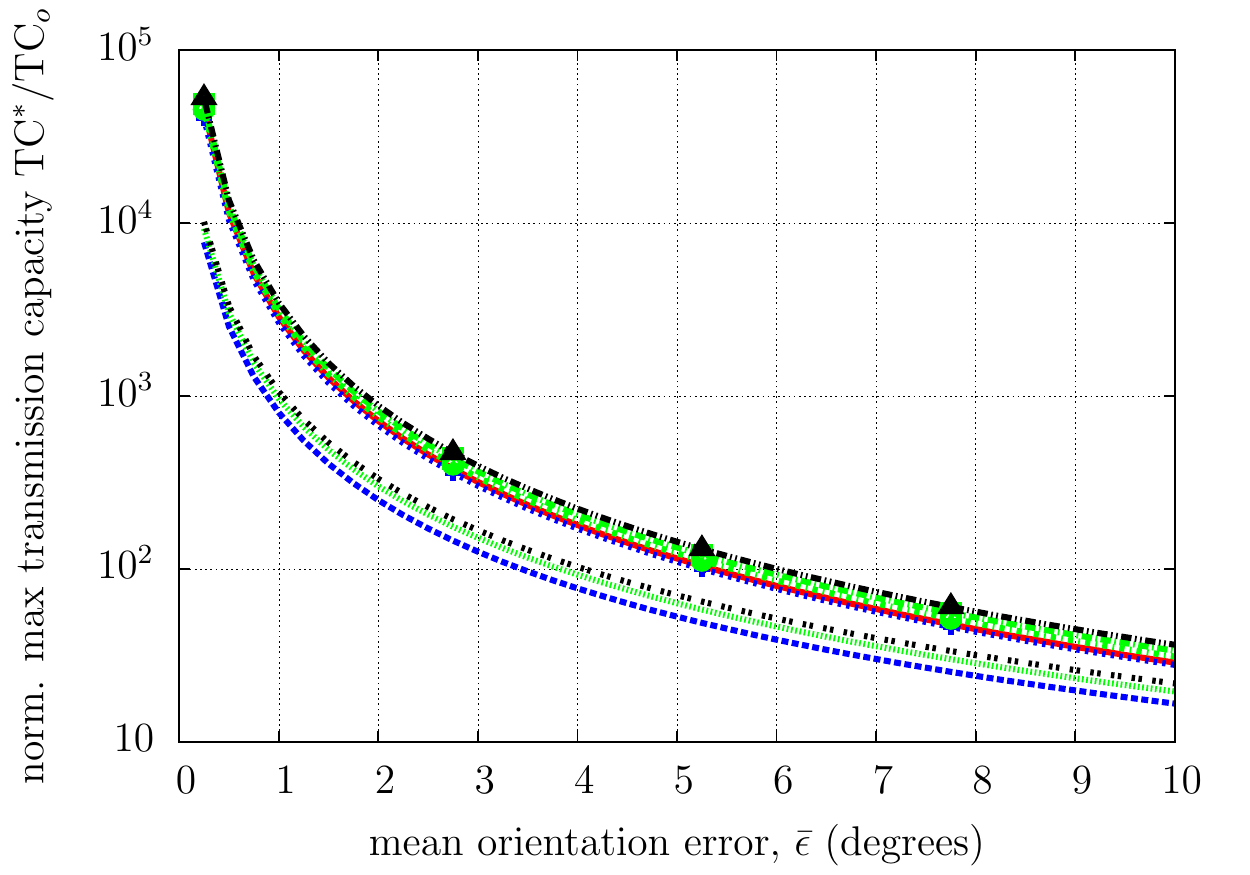}
    \end{subfigure}\\
    \begin{subfigure}[b]{0.4\linewidth}
      \centering
      \includegraphics[width=0.9\textwidth]{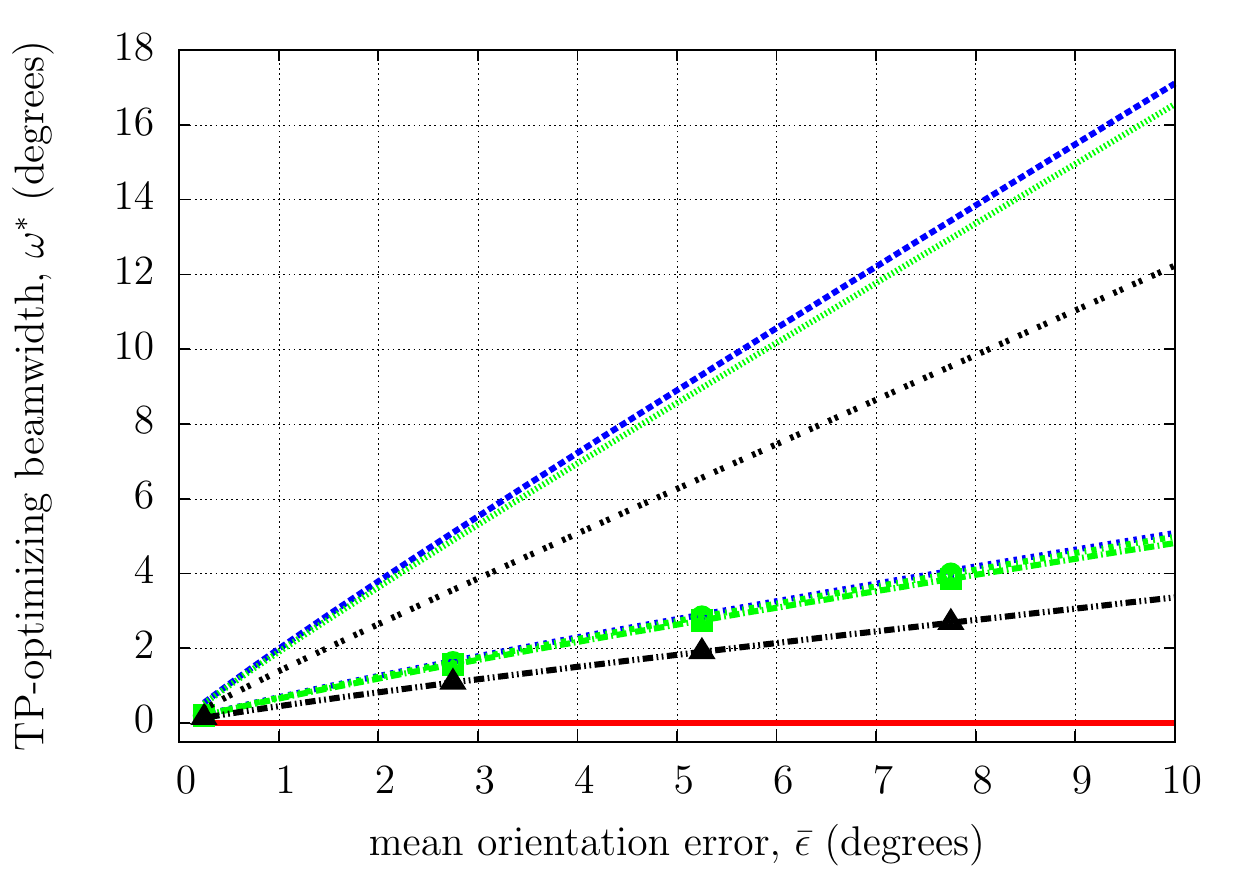}
    \end{subfigure}
    \begin{subfigure}[b]{0.4\linewidth}
      \centering
      \includegraphics[width=0.9\textwidth]{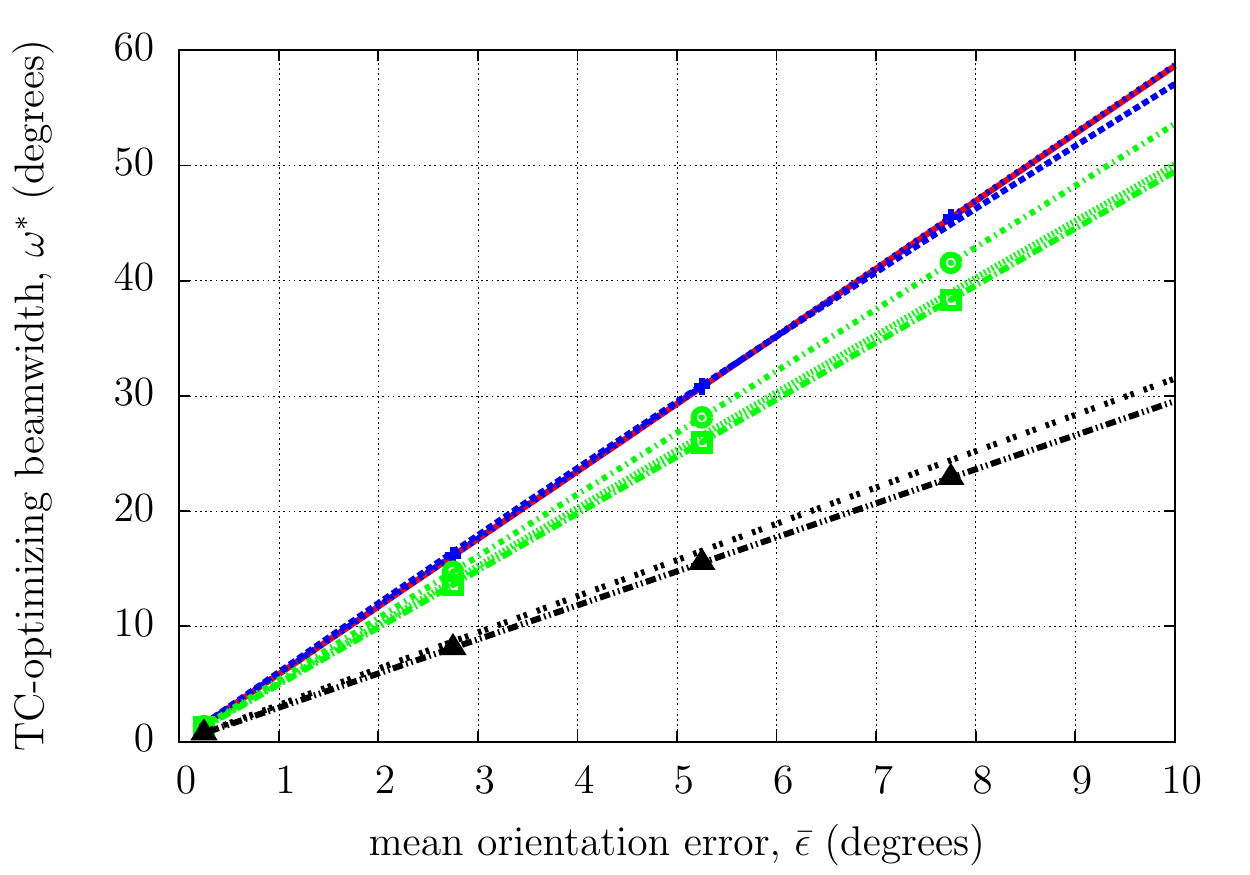}
    \end{subfigure}\\
    \begin{subfigure}[b]{0.8\linewidth}
      \centering
      \includegraphics[width=\textwidth]{legend}
    \end{subfigure}
  }{%
    \centering
    \begin{subfigure}[b]{0.5\linewidth}
      \centering
      \includegraphics[width=0.85\textwidth]{maxtp}
    \end{subfigure}%
    \begin{subfigure}[b]{0.5\linewidth}
      \centering
      \includegraphics[width=0.85\textwidth]{maxtc}
    \end{subfigure}\\
    \begin{subfigure}[b]{0.5\linewidth}
      \centering
      \includegraphics[width=0.85\textwidth]{omega_maxtp}
    \end{subfigure}%
    \begin{subfigure}[b]{0.5\linewidth}
      \centering
      \includegraphics[width=0.85\textwidth]{omega_maxtc}
    \end{subfigure}\\
    \begin{subfigure}[b]{\linewidth}
      \centering
      \includegraphics[width=0.9\textwidth]{legend}
    \end{subfigure}
  }
  \caption{%
    Plotted are spatial throughput \(\TP\) maximized over beamwidth (top-left), the resulting \(\TP\)-maximizing beamwidth (bottom-left), transmission capacity \(\TC\) maximized over beamwidth (top-right), and the resulting \(\TC\)-maximizing beamwidth (bottom-right).
    The legend, default parameters, and orientation error modeling are identical to that of \figref{numerical1}.
  }
  \label{fig:numerical2}
\end{figure*}

In \figref{numerical2}, we explore the maximization of \(\TP\) and \(\TC\) (\wrt{} beamwidth \(\omega\)) plotted against mean orientation error (top-left and top-right of \figref{numerical2}, respectively).
\correction{%
  In general, maximum spatial throughput \(\TP^*\) and maximum transmission capacity \(\TC^*\) decrease as the uncertainty in orientation increases and behave rather identically across radiation patterns.
  For the configurations plotted, the sidelobe strength \(g_2\) is the dominant factor in separating the throughput curves.
}

The corresponding maximizing beamwidths \(\omega^*\) for both maximum throughput metrics (bottom-left and bottom-right of \figref{numerical2}) are also plotted against mean orientation error.
Intuitively, the \(\TP\)- and \(\TC\)-maximizing beamwidths tend to increase with an increased uncertainty in antenna orientation.
\correction{
  Also, as expected under ideal sectors without sidelobes, \(\TP\)-monotonicity results in a maximizing beamwidth of zero regardless of the mean orientation error.
  The addition of sidelobes and transition widths into the sector pattern produces \(\TP\)-maximizing beamwidths larger than zero, and the outage-constrained nature of \(\TC^*\) produces significantly larger optimal beamwidths than \(\TP^*\).
  The sidelobe strength \(g_2\) seems to be the determining factor in grouping \(\TP\)-maximizing beamwidth curves, while the radiation pattern type (ideal, trans, 3GPP) appears to be the more dominant factor in grouping \(\TC\)-maximizing beamwidth curves.
  For the scenarios investigated, the \(\TC\)-maximizing beamwidth for the ideal sector without sidelobes seems to provide a good approximation for all other ideal sectors with sidelobes and also serves as an upper bound for the remaining patterns (trans and 3GPP).
}

\begin{remark}
  Interestingly, we see that both \(\TP\)- and \(\TC\)-maximizing beamwidths have a nearly linear relationship with the mean orientation error for all radiation patterns displayed.
  Assuming truncated exponential orientation error, the optimality constraint \eqref{tc-maximizer} can be reparameterized by the ratio \(\omega/\beps\), indicating that \(\omega^*\) does indeed scale linearly with \(\beps\) under the assumption of ideal sectors without sidelobes.
\end{remark}

%=======================================================================================================================
\section{Conclusions \& Future Work}\label{sec:conclusions}

In this paper, we introduced a model for capturing the effects of beam misdirection on coverage and throughput in a directional wireless network using stochastic geometry.
In networks employing ideal sector antennas without sidelobes, we found that the moderate assumption of a concave orientation error \cdf{} was sufficient to prove monotonicity and quasi-concavity (both with respect to antenna beamwidth) of spatial throughput (\(\TP\)) and outage-constrained transmission capacity (\(\TC\)), respectively.
Our numerical results confirm this, but also show that monotonicity of spatial throughput is not preserved for networks employing more complex antenna models.
However, unimodality appears to be maintained across the various radiation patterns studied for both throughput metrics, which warrants further investigation.

While varying the sector pattern's sidelobe strength and `sharpness' of the beamwidth, we found that the ideal sector pattern without sidelobes varied in its ability to approximate more complex patterns.
For instance, while the antenna sidelobe strength could greatly influence transmission capacity maximized over antenna beamwidth, the resulting maximizing beamwidths for different sidelobe strengths tended to be well approximated by that of the sector without sidelobes.
\correction{%
  There exist possible opportunities for upper bounding metrics (\ie{} transmission capacity-maximizing beamwidth) for complex radiation patterns by the use of simpler patterns (\ie{} ideal sector antenna without sidelobes).
}

Finally, we noted an apparent linear relationship between mean orientation error and throughput maximizing beamwidths.
This held across both throughput metrics and across the sector patterns explored in this paper, suggesting another interesting future direction of inquiry.

%%=======================================================================================================================
%\section{Future Work}\label{sec:future}
%
%\mytodo{add the future work section}

%\clearpage

\appendix

%=======================================================================================================================
\subsection{Proof of \prpref{success-general} (Success of a Typical Transmission)}\label{prf:success-general}

\begin{IEEEproof}
  A transmission is successful when the SINR is greater than or equal to \(\beta\):
  \ifdraft{%
    \begin{align}
      p_s &= \Pbb\left\{ \SINR_o \geq \beta \right\} \\
      &\overset{(a)}{=} \Pbb\left\{ H_{o,o} \geq \frac{\beta d^{\alpha}}{P_t\GT(\eps_{x_o})\GR(\eps_{y_o})}I_o \right\}
                     \Pbb\left\{H_{o,o} \geq \frac{\beta d^{\alpha} \eta}{P_t\GT(\eps_{x_o})\GR(\eps_{y_o})} \right\} \\
      &\overset{(b)}{=} \int_{0^+}^{\infty} \int_{0^+}^{\infty} \Ebb[e^{-sI_o}] e^{-\frac{\beta d^{\alpha} \eta}{P_t\gT\gR}} \fGT{\gT}\fGR{\gR}\drm\gT\drm\gR
    \end{align}
  }{%
    \begin{align}
      p_s &= \Pbb\left\{ \SINR_o \geq \beta \right\} \\
      &\overset{(a)}{=} \Pbb\left\{ H_{o,o} \geq \frac{\beta d^{\alpha}}{P_t\GT(\eps_{x_o})\GR(\eps_{y_o})}I_o \right\} * \nonumber\\
      &\qquad \qquad \Pbb\left\{H_{o,o} \geq \frac{\beta d^{\alpha} \eta}{P_t\GT(\eps_{x_o})\GR(\eps_{y_o})} \right\} \\
      &\overset{(b)}{=} \int_{0^+}^{\infty} \int_{0^+}^{\infty} \Ebb[e^{-sI_o}] e^{-\frac{\beta d^{\alpha} \eta}{P_t\gT\gR}} \fGT{\gT}\fGR{\gR}\drm\gT\drm\gR
    \end{align}
  }
  \begin{itemize}
    \item[(a)] expand \(\SINR_i \geq \beta\), isolate \(H_{o,o}\), and apply the memoryless property of \(H_{o,o}\), and
    \item[(b)] marginalize the gains between the typical TX and RX; the first term is the Laplace transform of the interference evaluated at \(s = \frac{\beta d^{\alpha}}{P_t\gT\gR}\); the second term is the \ccdf{} of \(H_{o,o}\).
  \end{itemize}

  Following Section 5.1.7 of \cite{Hae2013}, we work with \(\Ebb[e^{-sI_o}]\), an expectation over \(\Phi\), the fading variables \(H_{i,o}\), and the gains between interferers and the typical RX \(\{\GTI(\hat{\theta}_{x_i,y_o})\}\) and \(\{\GRI(\hat{\theta}_{y_o,x_i})\}\):
  \ifdraft{%
    \begin{align}
      \Lmc_I(s) &= \Ebb_{\hat{\Phi}}\left[ \prod_{x_i \in \Phi} e^{-sP_t \GTI(\hat{\theta}_{x_i,y_o})\GRI(\hat{\theta}_{y_o,x_i}) H_{i,o}d_{i,o}^{-\alpha}}\right] \\
      &\overset{(a)}{=} \Ebb_{\Phi}\left[ \prod_{x_i \in \Phi} \Ebb_{\GTI,\GRI,H}\left[ e^{-sP_t\GTI\GRI H d_{i,o}^{-\alpha}}\right]\right] \\
      &\overset{(b)}{=} \Ebb_{\Phi}\left[ \prod_{x_i \in \Phi} v(||x_j||) \right] \\
      &\overset{(c)}{=} \exp\left(-\int_{0}^{\infty} 1-v(x) \lambda(x)\drm x\right) \\
      &= \exp\left(-\int_{0}^{\infty} \Ebb_{\GTI,\GRI,H}\left[1-e^{-sP_t\GTI\GRI Hx^{-\alpha}}\right] \lambda(x)\drm x\right)
    \end{align}
  }{%
    \begin{align}
      \Lmc_I(s) &= \Ebb_{\hat{\Phi}}\left[ \prod_{x_i \in \Phi} e^{-sP_t \GTI(\hat{\theta}_{x_i,y_o})\GRI(\hat{\theta}_{y_o,x_i}) H_{i,o}d_{i,o}^{-\alpha}}\right] \\
      &\overset{(a)}{=} \Ebb_{\Phi}\left[ \prod_{x_i \in \Phi} \Ebb_{\GTI,\GRI,H}\left[ e^{-sP_t\GTI\GRI H d_{i,o}^{-\alpha}}\right]\right] \\
      &\overset{(b)}{=} \Ebb_{\Phi}\left[ \prod_{x_i \in \Phi} v(||x_j||) \right] \\
      &\overset{(c)}{=} e^{       -\int_{0}^{\infty} 1-v(x) \lambda(x)\drm x} \\
      &= e^{       -\int_{0}^{\infty} \Ebb_{\GTI,\GRI,H}\left[1-e^{-sP_t\GTI\GRI Hx^{-\alpha}}\right] \lambda(x)\drm x}
    \end{align}
  }
  \begin{itemize}
    \item[(a)] since \(x\), \(H\), \(\GTI\), and \(\GRI\) are all independent from each other, expectations are taken separately and brought into the product, while indexing on the fading and gains is dropped,
    \item[(b)] with \(y_o\) at the origin, we collapse \(\hat{\Phi}\) into a one-dimensional PPP on \(\Rbb^{+}\) with intensity \(\lambda(x) = \lambda 2\pi x, \forall x \in \Rbb^{+}\) and define \(v(x) = \Ebb_{\GTI,\GRI,H}\left[ e^{-sP_t\GTI\GRI Hx^{-\alpha}}\right]\), and
    \item[(c)] a mean of a product of \(v(x)\) over the collapsed \(\hat{\Phi}\) is a probability generating functional (pgfl) of the process; an explicit solution is given by Campbell's Theorem for PPPs.
  \end{itemize}

  Again mirroring the developments in \cite{Hae2013}, we work with the integral inside the exponential, \(x\), \(\GTI\), \(\GRI\), and \(H\) are independent, so their expectations can be evaluated separately:
  \begin{align}
    \int_{0}^{\infty} & \Ebb_{\GTI,\GRI,H}[1-e^{-sP_t\GTI\GRI Hx^{-\alpha}}] \lambda(x)\drm x \\
    \overset{(a)}{=}& \Ebb_{\GTI,\GRI,H}\left[ \int_{0}^{\infty} (1-e^{-sP_t\GTI\GRI Hx^{-\alpha}}) \lambda(x)\drm x \right] \\
    =& \Ebb_{\GTI,\GRI,H}\left[ \lambda\pi \int_{0}^{\infty} (1-e^{-sP_t\GTI\GRI Hx^{-\alpha}}) 2x \drm x \right] \\
    \overset{(b)}{=}& \Ebb_{\GTI,\GRI}\left[ \lambda\pi \Gamma(1+2/\alpha) \Gamma(1-2/\alpha) (sP_t\GTI\GRI)^{2/\alpha} \right] \\
    \overset{(c)}{=}& \lambda\pi \Gamma(1\!+\!2/\alpha)\Gamma(1\!-\!2/\alpha)(sP_t)^{2/\alpha} \Ebb\left[\GTI^{2/\alpha}\right] \Ebb\left[\GRI^{2/\alpha}\right]
  \end{align}
  \begin{itemize}
    \item[(a)] exchange the order of integration over \(\GTI\), \(\GRI\), and \(H\) with that of \(x\),
    \item[(b)] as done in Section 5.1.7 of \cite{Hae2013} in the case of Rayleigh fading with omni-directional antennas, the substitution of the integration variables, integration by parts, evaluation of the resulting integral, and finally taking the expectation over \(H\) produces:
      \ifdraft{%
        \begin{equation}
          \Ebb_{H}\left[ \lambda c_d \int_{0}^{\infty} (1-e^{-\hat{s}H x^{-\alpha}}) dx^{d-1} \drm x \right]
          = \lambda c_d \Gamma(1+\delta)\Gamma(1-\delta) \hat{s}^{\delta} \label{eq:haenggi}
        \end{equation}
      }{%
        \begin{align}
          \Ebb_{H}\left[ \lambda c_d \int_{0}^{\infty} (1-e^{-\hat{s}H x^{-\alpha}}) dx^{d-1} \drm x \right] \nonumber\\
          = \lambda c_d \Gamma(1+\delta)\Gamma(1-\delta) \hat{s}^{\delta} \label{eq:haenggi}
        \end{align}
      }
      where \(d\) is the dimension of the space in which the points of \(\Phi\) reside, \(\delta = d/\alpha\) and \(c_d\) is a constant depending on \(d\).

      In our case, \(d = 2\) and it follows that \(\delta = 2/\alpha\) and \(c_2 = \pi\).
      By substituting \(\hat{s} = sP_t\GTI\GRI\) and taking the expectation of both sides of \eqref{haenggi} \wrt{} \(\GTI\) and \(\GRI\), we complete this step.
    \item[(c)] expectations over \(\GTI\) and \(\GRI\) reduce to taking moments of each \rv{}, (as noted in \cite{WanRee2012}).
  \end{itemize}

  Now, the Laplace transform can be expressed:
  \ifdraft{%
    \begin{equation}
      \Lmc_I(s) = \Ebb\left[e^{-sI_o}\right]
      = \exp\left(- \lambda\pi \Gamma(1+2/\alpha)\Gamma(1-2/\alpha)(sP_t)^{2/\alpha} \Ebb\left[\GTI^{2/\alpha}\right] \Ebb\left[\GRI^{2/\alpha}\right]\right).
    \end{equation}
  }{%
    \begin{align}
      \Lmc_I(s) &= \Ebb\left[e^{-sI_o}\right] \\
      &= e^{      - \lambda\pi \Gamma(1+2/\alpha)\Gamma(1-2/\alpha)(sP_t)^{2/\alpha} \Ebb\left[\GTI^{2/\alpha}\right] \Ebb\left[\GRI^{2/\alpha}\right]}.
    \end{align}
  }
  With \(s = \frac{\beta d^{\alpha}}{P_t\gT\gR}\), we express success probability of a transmission between the typical pair \(o\):
  \ifdraft{%
    \begin{align}
      p_s &= \int_{0^+}^{\infty} \!\! \int_{0^+}^{\infty} \Ebb[e^{-sI_o}] e^{-\frac{\beta d^{\alpha} \eta}{P_t\gT\gR}} \fGT{\gT}\fGR{\gR}\drm\gT\drm\gR \\
      &= \int_{0^+}^{\infty} \!\! \int_{0^+}^{\infty} \exp\left(-\lambda \pi \Gamma(1+2/\alpha)\Gamma(1-2/\alpha)(sP_t)^{2/\alpha} \Ebb\left[\GTI^{2/\alpha}\right] \Ebb\left[\GRI^{2/\alpha}\right]\right) * \nonumber\\
      &\qquad \qquad e^{-\frac{\beta d^{\alpha} \eta}{P_t\gT\gR}} \fGT{\gT}\fGR{\gR}\drm\gT\drm\gR. \IEEEQEDhere
    \end{align}
  }{%
    \begin{align}
      p_s &= \int_{0^+}^{\infty} \int_{0^+}^{\infty} \Ebb[e^{-sI_o}] e^{-\frac{\beta d^{\alpha} \eta}{P_t\gT\gR}} \fGT{\gT}\fGR{\gR}\drm\gT\drm\gR \\
      &= \int_{0^+}^{\infty}      \int_{0^+}^{\infty} e^{       -\lambda \pi \Gamma(1+2/\alpha)\Gamma(1-2/\alpha)(sP_t)^{2/\alpha} \Ebb\left[\GTI^{2/\alpha}\right] \Ebb\left[\GRI^{2/\alpha}\right]} * \nonumber\\
      &\qquad \qquad e^{-\frac{\beta d^{\alpha} \eta}{P_t\gT\gR}} \fGT{\gT}\fGR{\gR}\drm\gT\drm\gR.
    \end{align}
  }
\end{IEEEproof}

%=======================================================================================================================
\subsection{Proof of \prpref{tp-sector-noside} (\(\TP\) using Sectors \correction{without} Sidelobes)}\label{prf:tp-sector-noside}

\begin{IEEEproof}
  The first and second derivatives of \(\lambda p_s(\lambda)\) are:
  \ifdraft{%
    \begin{align}
      \frac{\drm}{\drm\lambda}     \lambda p_s(\lambda) &= (1-A\lambda)Ce^{-\lambda A - B},
    & \frac{\drm^2}{\drm\lambda^2} \lambda p_s(\lambda) &= (A\lambda-2)ACe^{-\lambda A - B},
    \end{align}
  }{%
    \begin{align}
      \frac{\drm}{\drm\lambda}     \lambda p_s(\lambda) &= (1-A\lambda)Ce^{-\lambda A - B} \\
      \frac{\drm^2}{\drm\lambda^2} \lambda p_s(\lambda) &= (A\lambda-2)ACe^{-\lambda A - B},
    \end{align}
  }
  where \(A = \pi\kappa d^2\beta^{2/\alpha} p^2\), \(B = \frac{\beta d^\alpha \eta}{P_t g_1^2}\), and \(C = u^2\).
  A single root of the first derivative exists at \(\lambda^* = 1/A\), while the second derivative, when evaluated at \(\lambda^*\), is negative:
  \begin{equation}
    \left.\frac{\drm^2}{\drm\lambda^2} \lambda p_s(\lambda)\right|_{\lambda=\lambda^*} = -ACe^{-1-B} \leq 0,
  \end{equation}
  due to \(A > 0\), \(e^{-1-B} > 0\), and \(C = u^2 = \sqFeps{\frac{\omega}{2}} > 0\) when \(\omega > 0\).
  Thus, due to the first and second derivative tests, we have that \(\lambda^*\) is the global maximizer of \(\lambda p_s(\lambda)\).
\end{IEEEproof}

%=======================================================================================================================
\subsection{Proof of \prpref{tp-monotone} (Concave \({\protect\Feps{}}\) Implies Monotonicity of \(\TPs\) in Beamwidth)}\label{prf:tp-monotone}

\begin{IEEEproof}
  We rewrite spatial throughput \eqref{tp-sector-noside} by expanding \(u\), \(p\), \(g_1\) in terms of \(x = \omega/2\) and study \(\TPs(x)\) over \(x \in [0,\pi]\).
  Specifically, we show that \(\TPs(x)\) is monotone decreasing over \((0,\pi]\).
  To do so, we will need \(\TPs\) and its derivative \wrt{} \(x\):
  \begin{align}
    \TPs(x)   &= \frac{\sqFeps{x} e^{-Bx^2}}{Ax^2} \\
    \TPs'(x)  &= -\frac{2e^{-Bx^2}\Feps{x}}{Ax^3}\left((1+Bx^2)\Feps{x} - x\feps{x}\right),
  \end{align}
  with non-negative constants \(A = \frac{1}{\pi}e \kappa d^2 \beta^{2/\alpha}\) and \(B = \frac{\beta d^\alpha \eta}{\pi^2 P_t}\).
  Since \(\frac{2e^{-Bx^2}\Feps{x}}{Ax^3} > 0\) for all \(x \in (0,\pi]\), it suffices to show:
  \begin{align}
    0 &\leq \left((1+Bx^2)\Feps{x} - x\feps{x}\right) \\
    \frac{\feps{x}}{\Feps{x}} &\leq \frac{1}{x}+Bx, \quad \forall x \in (0,\pi].
  \end{align}
  in order to prove spatial throughput is monotone decreasing in \(x\), \(\TPs'(x) \leq 0\).

  By the assumption of concavity over \([0,\pi]\), \(\Feps{}\) evaluated at \(y \in [0,\pi]\) lies below its first order Taylor series approximation centered at \(x \in [0,\pi]\):
  \begin{equation}
    \Feps{y} \leq \Feps{x} + \feps{x}(y-x), \quad \forall x,y \in [0,\pi].
  \end{equation}
  After setting \(y = 0\), \(\Feps{0} = 0\), rearranging the result, and adding a positive quantity \(Bx^2\) to the \rhs{}, we can conclude our proof:
  \begin{equation}
    \frac{\feps{x}}{\Feps{x}} \leq \frac{1}{x} < \frac{1}{x} + Bx^2, \quad \forall x \in (0,\pi]. \IEEEQEDhereeqn\
  \end{equation}
\end{IEEEproof}

%=======================================================================================================================
\subsection{Proof of \corref{tp-omni} (\(\TP\) using Omni-directional Antennas)}\label{prf:tp-omni}

\begin{IEEEproof}
  The proof of \prpref{tp-sector-noside} can be used with \(A = \pi\kappa d^2\beta^{2/\alpha}\), \(B = \frac{\beta d^\alpha \eta}{P_t}\), and \(C = 1\).
  Since \(A\), \(e^{-B}\), and \(C\) are all positive, \(\lambda^* = 1/A\) is the global maximizer of \(\lambda p_s(\lambda)\).
\end{IEEEproof}

%=======================================================================================================================
\subsection{Proof of \prpref{tc-sector-noside} (\(TC\) with Sectors \correction{without} Sidelobes)}\label{prf:tc-sector-noside}

\begin{IEEEproof}
  Rewrite \eqref{success-sector-noside} as \(p_s = C e^{-\lambda A - B}\), where \(A = \pi\kappa d^2\beta^{2/\alpha} p^2\), \(B = \frac{\beta d^\alpha \eta}{P_t g_1^2}\), and \(C = u^2\).
  Solving for \(\lambda\) yields \(\lambda(p_e) = \log\left(\frac{Ce^{-B}}{(1-p_e)}\right)/A\).
  %
  %\begin{equation}
  %  \lambda = \frac{\log\left(\frac{Ce^{-B}}{(1-p_e)}\right)}{A}.
  %\end{equation}
  %
  Note that \(e^{-B} = 1-(1-e^{-B})\), and let \(p_{\eta} = 1-e^{-B}\), which represents fading outage due to background noise.
  Multiplying \(\lambda(p_e)\) by the success rate (\(1-p_e\)) provides the maximum intensity of successful transmissions, subject to outage \(p_e\).
\end{IEEEproof}

%=======================================================================================================================
\subsection{Proof of \prpref{tc-unimodal} (Concavity of \({\protect\Feps{\cdot}}\) Implies Unimodality of \(\TCs\))}\label{prf:tc-unimodal}

\begin{IEEEproof}
  We rewrite transmission capacity \eqref{tc-sector-noside} by expanding \(u\), \(p\), \(g_1\) in terms of \(x = \omega/2\) and study \(\TCs(x)\) over \(x \in (0,\pi]\).
  Specifically, we show that \(\TCs(x)\) is \emph{i)} monotone increasing over \((0,x_l]\), \emph{ii)} quasiconcave over \([x_l,x_u]\), and \emph{iii)} monotone decreasing over \((x_u,\pi]\), where \(x_l = \invFeps{\sqrt{1-p_e}}\) and \(x_u = \epsmax\).
  Once these three facts are established and combined with the continuity of \(\TCs\) over \((0,\pi]\), we can readily conclude that the unique maximizer of \(\TCs\) lies between \((x_l,x_u]\) and that \(\TCs\) is quasiconcave (unimodal) over this domain.
  To do so, we will need \(\TCs\) and its first two derivatives \wrt{} \(x\):
  \ifdraft{%
    \begin{align}
      \TCs(x)   &= \frac{ A}{x^2} \left( 2\log(F(x)) + B \right) - AC \label{eq:tc-d0} \\
      \TCs'(x)  &= \frac{2A}{x^2} \left( \frac{f(x)}{F(x)}-\frac{2\log(F(x)) + B}{x} \right) \label{eq:tc-d1} \\
      \TCs''(x) &= \frac{2A}{x^3}
                    \left( 3\frac{2\log(F(x)) + B}{x}-\frac{4 f(x)}{F(x)} - \frac{xf(x)^2}{F(x)^2} + \frac{xf'(x)}{F(x)}\right), \label{eq:tc-d2}
    \end{align}
  }{%
    \begin{align}
      \TCs(x)   &= \frac{ A}{x^2} \left( 2\log(F(x)) + B \right) - AC \label{eq:tc-d0} \\
      \TCs'(x)  &= \frac{2A}{x^2} \left( \frac{f(x)}{F(x)}-\frac{2\log(F(x)) + B}{x} \right) \label{eq:tc-d1} \\
      \TCs''(x) &= \frac{2A}{x^3} * \nonumber\\
      &\hspace{-1em}\left( 3\frac{2\log(F(x)) + B}{x}-\frac{4 f(x)}{F(x)} - \frac{xf(x)^2}{F(x)^2} + \frac{xf'(x)}{F(x)}\right), \label{eq:tc-d2}
    \end{align}
  }
  with positive constants \(A = \frac{\pi(1-p_e)}{\kappa d^2\beta^{2/\alpha}}\), \(B = \log\left(\frac{1}{1-p_e}\right)\), and \(C = \frac{\beta d^\alpha \eta}{P_t \pi^2}\).
  Note: \(\TCs\) is smooth at \(x_l\) and \(\TCs'(x_l) > 0\), but may not be differentiable (\ie{} have a sharp turn) at \(x_u\).

  For \emph{i)}, note that \(2\log(F(x)) + B\) is monotone increasing in \(x\) due to the monotonicity of \(F\) and \(\log\).
  It follows that when \(x \leq x_l\):
  \begin{align}
    2\log(F(x)) + B &\leq 2\log(F(x_l)) + B \\
    &\leq 2\log\left(F\left(\invFeps{\sqrt{1-p_e}}\right)\right) + B \\
    &= \log(1-p_e) + \log(1/(1-p_e)) = 0.
  \end{align}
  Substituting this bound into \eqref{tc-d1}, we obtain the desired monotonicity of \(\TCs\):
  \begin{equation}
    \TCs'(x) \geq \frac{2A}{x^2} \frac{f(x)}{F(x)} > 0, \quad \forall x \in (0,x_l).
  \end{equation}

  For \emph{ii)}, we use a sufficient condition for quasiconcavity from Boyd and Vandenberghe \cite{BoyVan2004}:
  \begin{equation}
    \TCs'(x) = 0 \Rightarrow \TCs''(x) < 0, \quad \forall x \in (x_l,x_u).
  \end{equation}
  In words, if all stationary points are associated with local maxima, then only a single stationary point exists, which necessarily provides the global maximum.
  Let \(x^* \in (x_l,x_u)\) be a stationary point of \(\TCs\).  From \eqref{tc-d1}, \(\TCs(x^*) = 0\) implies:
  \begin{equation}\label{eq:tc-stationary}
    \frac{2\log(F(x^*))+B}{x^*} = \frac{f(x^*)}{F(x^*)}.
  \end{equation}
  Simplify \eqref{tc-d2} at this stationary point by substitution of the above equality:
  \ifdraft{%
    \begin{equation}
      \TCs''(x^*) = \frac{2A}{(x^*)^3}
      \left( -\frac{f(x^*)}{F(x^*)} - \frac{x^*f(x^*)^2}{F(x^*)^2} + \frac{x^*f'(x^*)}{F(x^*)}\right).
    \end{equation}
  }{%
    \begin{align}
      \TCs''(x^*) &= \frac{2A}{(x^*)^3} * \\
    & \left( -\frac{f(x^*)}{F(x^*)} - \frac{x^*f(x^*)^2}{F(x^*)^2} + \frac{x^*f'(x^*)}{F(x^*)}\right).
    \end{align}
  }
  Over \((x_l,x_u)\), we have \(F(x) > 0\), \(f(x) > 0\), and \(f'(x) \leq 0\), thus \(\TCs''(x^*) < 0\) and \(\TCs\) is unimodal over \((x_l,x_u)\) and thus \([x_l,x_u]\) by continuity of \(\TCs\).

  For \emph{iii)}, note that \(F(x) = 1\) and \(f(x) = 0\) for all \(x > x_u\).  Thus, \eqref{tc-d1} can be simplified:
  \begin{equation}
    \TCs'(x) = \frac{-2AB}{x^3} < 0, \quad \forall x \in (x_u,\pi].
  \end{equation}

  Finally, since \(\TCs\) is increasing on \((0,x_l]\) and decreasing on \((x_u,\pi]\), the maximization of \(\TCs(x)\) can be reduced to searching over the remaining unimodal portion of \(\TCs\): \((x_l,x_u]\).
\end{IEEEproof}

%=======================================================================================================================
\subsection{Proof of \corref{tc-maximizer} (Conditions on the Maximizing \(\omega^*\) for \(\TCs\))}\label{prf:tc-maximizer}

\begin{IEEEproof}
  First, by \prpref{tc-unimodal} and its proof in \prfref{tc-unimodal}, we know that \(\TCs(x)\) with \(x = \omega/2\) is unimodal and contains a unique maximizer within: \(x^* \in \left(\invFeps{\sqrt{1-p_e}},\epsmax\right]\).
%  \begin{equation}
%    x^* \in \left(\invFeps{\sqrt{1-p_e}},\epsmax\right].
%  \end{equation}

  \(\TCs(x)\) may have a sharp turn at \(x = \epsmax\), so we take the left derivative of \(\TCs\) using \eqref{tc-d1}.
  Since \(F(\epsmax) = 1\), we have:
  \ifdraft{%
    \begin{equation}
      \TCs'(\epsmax^{-}) = \lim_{x\rightarrow\epsmax^{-}} \TCs'(x)
      = \frac{2A}{\epsmax^2} \left( f(\epsmax) - \frac{B}{\epsmax} \right),
    \end{equation}
  }{%
    \begin{align}
      \TCs'(\epsmax^{-}) &= \lim_{x\rightarrow\epsmax^{-}} \TCs'(x) \\
    & = \frac{2A}{\epsmax^2} \left( f(\epsmax) - \frac{B}{\epsmax} \right),
    \end{align}
  }
  with positive constants \(A = \frac{\pi(1-p_e)}{\kappa d^2\beta^{2/\alpha}}\), \(B = \log\left(\frac{1}{1-p_e}\right)\).

  First, when \(f(\epsmax) < \frac{B}{\epsmax}\), we have \(\TCs'(\epsmax^{-}) < 0\) and the maximizing \(x^*\) must lie strictly less than \(\epsmax\).
  Second, when \(f(\epsmax) > \frac{B}{\epsmax}\), we have \(\TCs'(\epsmax^{-}) > 0\) and the maximizing \(x^*\) must be exactly \(\epsmax\) due to the unimodality of \(\TCs\) over the rest of \(\left(\invFeps{\sqrt{1-p_e}},\epsmax\right]\).
  Lastly, when \(f(\epsmax) = \frac{B}{\epsmax}\), we have \(\TCs'(\epsmax^{-}) = 0\).
  We then take the left second derivative of \(\TCs\) at \(\epsmax\) using \eqref{tc-d2}, and since \(\epsmax\) is a stationary point of \(\TCs\), we can apply \eqref{tc-stationary}:
  \ifdraft{%
    \begin{align}
      \TCs''&(\epsmax^{-}) = \lim_{x\rightarrow\epsmax^{-}} \TCs''(x) \\
      &= \frac{2A}{\epsmax^3} \left( 3\frac{2\log(F(\epsmax)) + B}{\epsmax}-\frac{4 f(\epsmax)}{F(\epsmax)}
                   - \frac{\epsmax f(\epsmax)^2}{F(\epsmax)^2} + \frac{\epsmax f'(\epsmax^{-})}{F(\epsmax)}\right) \\
      &= \frac{2A}{\epsmax^3} \left( -\frac{f(\epsmax)}{F(\epsmax)} - \frac{\epsmax f(\epsmax)^2}{F(\epsmax)^2} + \frac{\epsmax f'(\epsmax^{-})}{F(\epsmax)}\right)\\
      &= \frac{2A}{\epsmax^3} \left( -f(\epsmax) - \epsmax f(\epsmax)^2 + \epsmax f'(\epsmax^{-}) \right)
    \end{align}
  }{%
    \begin{align}
      \TCs''&(\epsmax^{-}) = \lim_{x\rightarrow\epsmax^{-}} \TCs''(x) \\
      &= \frac{2A}{\epsmax^3} \left( 3\frac{2\log(F(\epsmax)) + B}{\epsmax}-\frac{4 f(\epsmax)}{F(\epsmax)} \right.\\
      &\qquad\left.- \frac{\epsmax f(\epsmax)^2}{F(\epsmax)^2} + \frac{\epsmax f'(\epsmax^{-})}{F(\epsmax)}\right) \\
      &= \frac{2A}{\epsmax^3} \left( -\frac{f(\epsmax)}{F(\epsmax)} - \frac{\epsmax f(\epsmax)^2}{F(\epsmax)^2} + \frac{\epsmax f'(\epsmax^{-})}{F(\epsmax)}\right)\\
      &= \frac{2A}{\epsmax^3} \left( -f(\epsmax) - \epsmax f(\epsmax)^2 + \epsmax f'(\epsmax^{-}) \right)
    \end{align}
  }
  where \(F(\epsmax) = 1\).
  Since \(\epsmax > 0\), \(f(\epsmax) > 0\), and \(f'(\epsmax^{-}) \leq 0\), we conclude that \(\TCs''(\epsmax^{-}) < 0\) and \(\TCs\) is concave down at \(\epsmax^{-}\).
  Thus, the maximizing \(x^*\) must be exactly \(\epsmax\) due to the unimodality of \(\TCs\) over the rest of \(\left(\invFeps{\sqrt{1-p_e}},\epsmax\right]\).
\end{IEEEproof}

%=======================================================================================================================
\subsection{Proof of \corref{tc-omni} (\(TC\) with Omni-directional Antennas)}\label{prf:tc-omni}

\begin{IEEEproof}
  The proof of \prpref{tc-sector-noside} can be used with \(A = \pi\kappa d^2\beta^{2/\alpha}\), \(B = \frac{\beta d^\alpha \eta}{P_t}\), and \(C = 1\).
\end{IEEEproof}

%% use section* for acknowledgement
%\section*{Acknowledgment}

\ifCLASSOPTIONcaptionsoff
\newpage
\fi



\bibliographystyle{IEEEtran}
\bibliography{IEEEabrv,refs}

%\begin{thebibliography}{1}
%
%    \bibitem{IEEEhowto:kopka}
%    H.~Kopka and P.~W. Daly, \emph{A Guide to \LaTeX}, 3rd~ed.\hskip 1em plus
%    0.5em minus 0.4em\relax Harlow, England: Addison-Wesley, 1999.
%
%\end{thebibliography}


\ifdraft{%
  % no bios
}{%
  \begin{IEEEbiography}[{\includegraphics[width=1in,height=1.25in,clip,keepaspectratio]{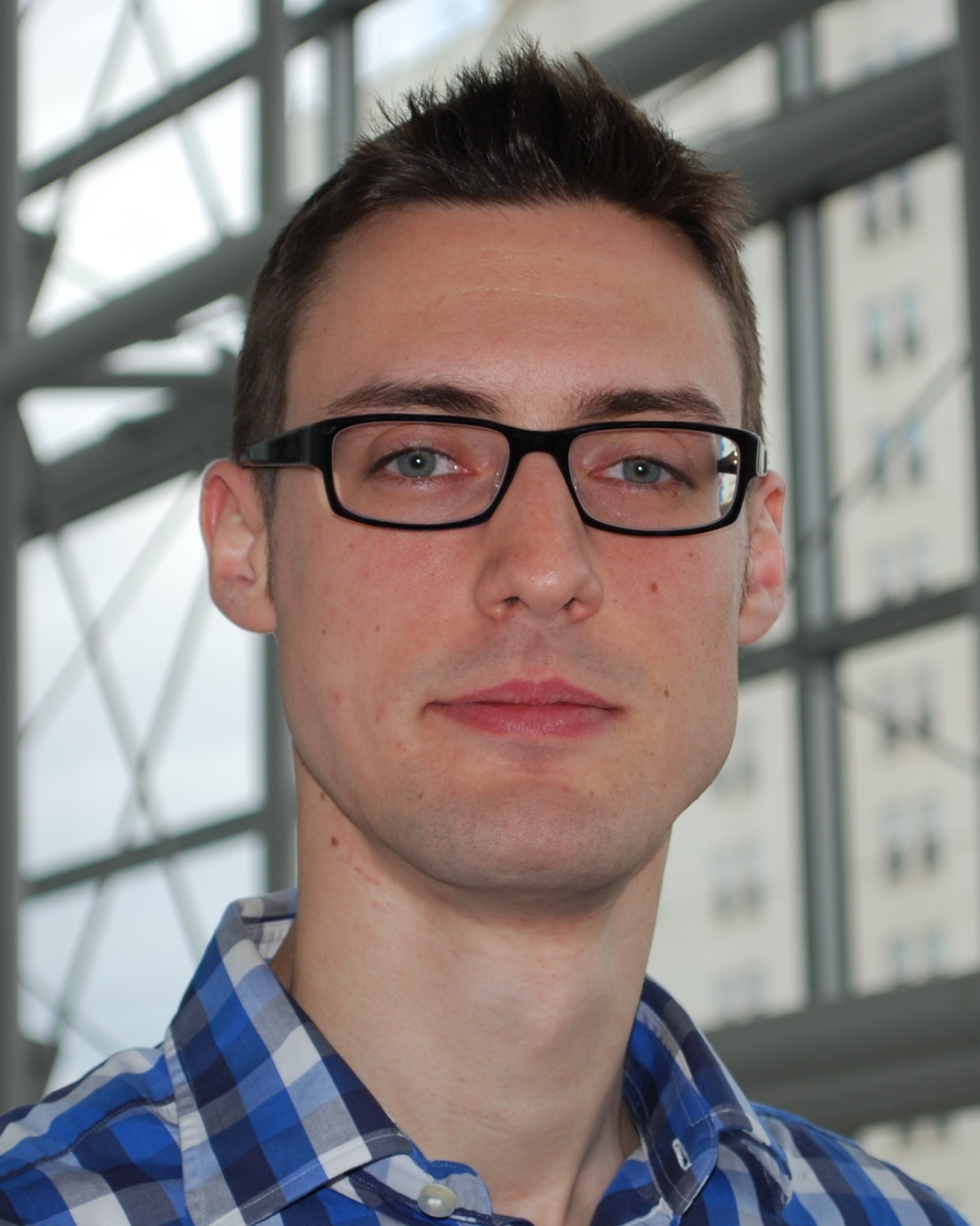}}]{Jeffrey Wildman}%
    received his dual M.S./B.S. degree in electrical engineering from Drexel University in 2009.
    He has interned and visited with the Wideband Tactical Networking Group at MIT Lincoln Laboratory (2007, 2009), with the Network Operating Systems Technology Group at Cisco Systems, Inc. (2012), and most recently with the Center for Wireless Communications at the University of Oulu, Finland (2013).
    Jeffrey is currently pursuing a Ph.D. in electrical engineering at Drexel University with interests in the modeling and analysis of wireless communications systems.
  \end{IEEEbiography}

  \vspace*{-2\baselineskip}

  \begin{IEEEbiography}[{\includegraphics[width=1in,height=1.25in,clip,keepaspectratio]{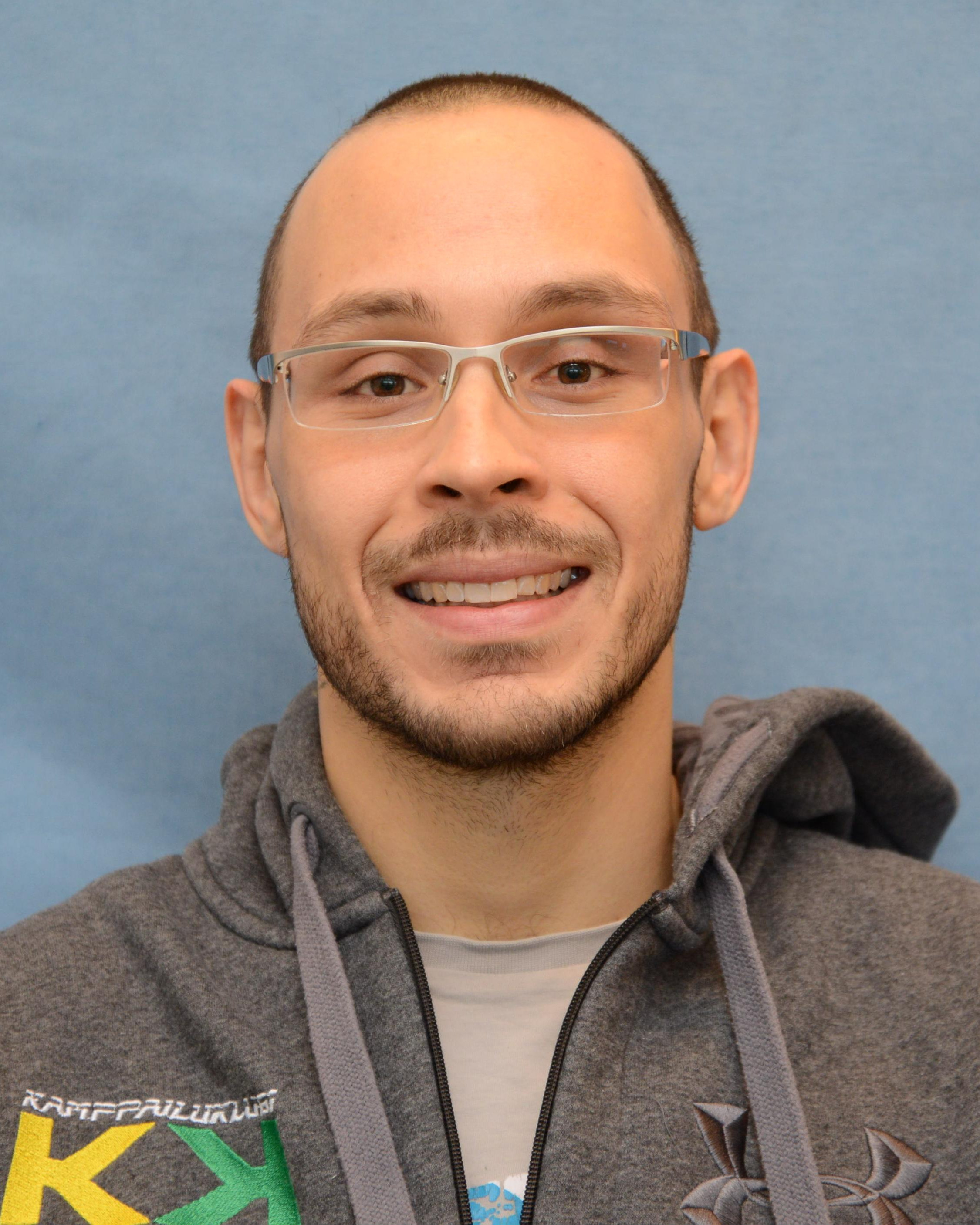}}]{Pedro Henrique Juliano Nardelli}%
    received the B.S. and M.Sc. degrees in electrical engineering from the State University of Campinas, Brazil, in 2006 and 2008, respectively.
    In 2013 he received his doctoral degree from University of Oulu, Finland, and State University of Campinas following a dual-degree agreement.
    Nowadays he holds a postdoctoral position at University of Oulu, and his studies are mainly focused on the efficiency of wireless networks and spatio-temporal dynamics of complex systems.
  \end{IEEEbiography}

  \vspace*{-2\baselineskip}

  \begin{IEEEbiography}[{\includegraphics[width=1in,height=1.25in,clip,keepaspectratio]{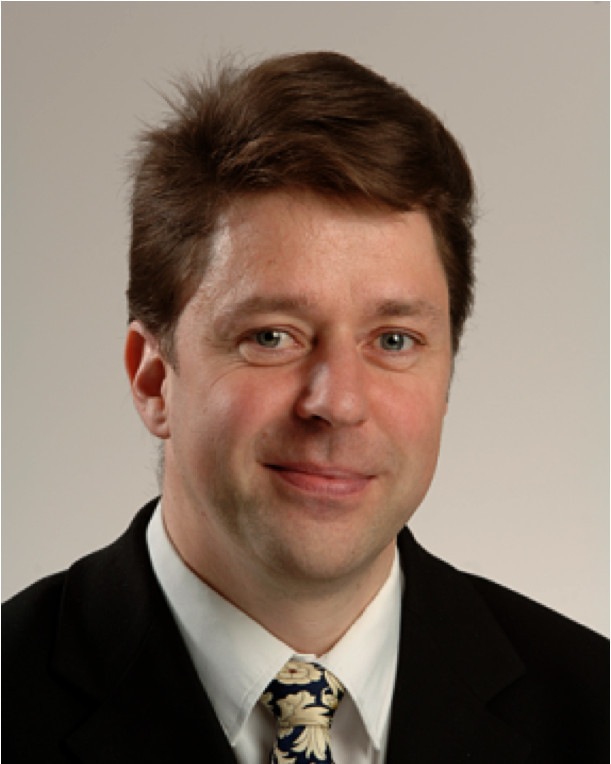}}]{Matti Latva-aho}%
    was born in Kuivaniemi, Finland in 1968.
    He received the M.Sc., Lic.Tech. and Dr. Tech (Hons.) degrees in Electrical Engineering from the University of Oulu, Finland in 1992, 1996 and 1998, respectively.
    From 1992 to 1993, he was a Research Engineer at Nokia Mobile Phones, Oulu, Finland.  During the years 1994-1998 he was a Research Scientist at Telecommunication Laboratory and Centre for Wireless Communications at the University of Oulu.
    Currently he is the Department Chair Professor of Digital Transmission Techniques and Head of Department at the University of Oulu, Department for Communications Engineering.
    Prof. Latva-aho was Director of Centre for Wireless Communications at the University of Oulu during the years 1998-2006.
    His research interests are related to mobile broadband wireless communication systems.
    Prof. Latva-aho has published over 200 conference or journal papers in the field of wireless communications.
    He has been TPC Chairman for PIMRC’06, TPC Co-Chairman for ChinaCom’07 and General Chairman for WPMC’08.
    He acted as the Chairman and vice-chairman of IEEE Communications Finland Chapter in 2000-2003.
  \end{IEEEbiography}

  \vspace*{-2\baselineskip}

  \begin{IEEEbiography}[{\includegraphics[width=1in,height=1.25in,clip,keepaspectratio]{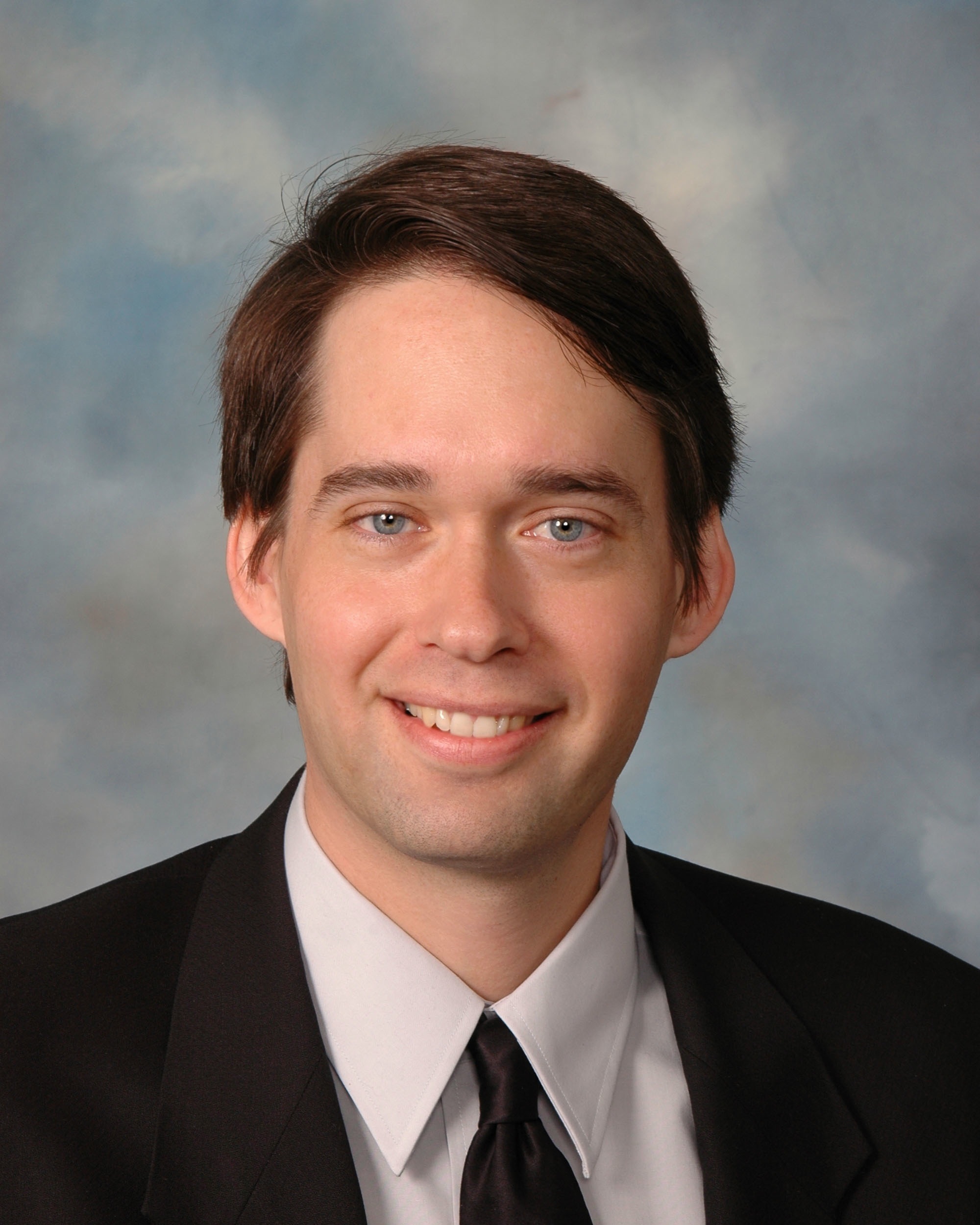}}]{Steven Weber}%
    (SM 1997, M 2003, S 2011) received his B.S. degree in 1996 from Marquette University in Milwaukee, WI, and his M.S. and Ph.D. degrees from The University of Texas at Austin in 1999 and 2003 respectively.
    He joined the Department of Electrical and Computer Engineering at Drexel University in 2003 where he is currently an associate professor.
    His research interests are centered around mathematical modeling of computer and communication networks, specifically streaming multimedia and ad hoc networks.
  \end{IEEEbiography}
}

\end{document}